\documentclass[aps,pra,preprint,superscriptaddress,longbibliography]{revtex4-2}

\usepackage{amsmath}
\usepackage{amsthm}
\usepackage{amsfonts}
\usepackage{amssymb}
\usepackage{xcolor,graphicx}
\usepackage{tikz}
\usepackage{color}
\usepackage[pdftex]{hyperref} 
\usepackage{longtable}
\usepackage{array}
\usepackage{dsfont}

\newcommand{\vect}[1]{\ensuremath{\mathbf #1}}

\begin{document}
	
	\title{Spatial light mode analogues of generalized quantum coherent states}

        \author{M.~P. Morales Rodr\'iguez}
        \affiliation{Departamento de F\'sica Te\'orica, At\'omica y \'Optica, Universidad de Valladolid, Valladolid, 47011 Spain.}

        \author{E. Garc\'ia Herrera}

        \author{O. Magaña Loaiza}
        \email[e-mail: ]{maganaloaiza@lsu.edu}
        \affiliation{Louisiana State University, Department of Physics \& Astronomy, Baton Rouge, LA 70803-4001}

        \author{B. Perez-Garcia}
        \email[e-mail: ]{b.pegar@tec.mx}
        \affiliation{Photonics and Mathematical Optics Group, Tecnologico de Monterrey, Monterrey 64849, Mexico} 

        \author{F. Marroqu\'in Guti\'errez}
        \email[e-mail: ]{fco.marroquin.gtz@upp.edu.mx }
        
        \author{B.~M. Rodr\'iguez-Lara}
        \email[e-mail: ]{bmlara@upp.edu.mx; blas.rodriguez@gmail.com}
        \affiliation{Universidad Polit\'ecnica de Pachuca. Carr. Pachuca-Cd. Sahag\'un Km.20, Ex-Hda. Santa B\'arbara. Zempoala, 43830 Hidalgo, Mexico }
	
	\date{\today}

 \begin{abstract}
 We use the spatial degree of freedom of light modes to construct optical analogues of generalized quantum coherent states for Hermite- and Laguerre-Gauss modes. 
Our optical analogues preserve the statistical properties of their quantum counterparts, encoded in their amplitude and phase distributions.  
We explore three basic symmetries that provide generalized displaced, rotated, and squeezed coherent states. 
Given the substantial interest in squeezed states for probing matter, we believe that the optical analogues introduced here have significant implications for optical sensing. 
Specifically, the single-particle nature of our spatial modes makes them robust candidates for sensing photosensitive materials. 
Overall, our approach opens the door to optical metrology and sensing protocols that mimic those already existing in the quantum realm, and facilitates further exploration of the quantum state zoo through classical optical analogues.
 \end{abstract}

\maketitle
\newpage

\section{Introduction}

The interest in classical optical analogues of quantum systems arises from a compelling reason. 
The theoretical formalism that describes the degrees of freedom of light modes is mathematically similar to those for quantum multiparticle systems \cite{Hashemi2015,Qian2015,Ndagano2017,Fabre2020,Shen2022}. 
This convergence highlights the potential of classical optical systems to mimic quantum behavior \cite{Longhi2009,Park2012,Dragoman2013}. 
They also offer deeper insights into fundamental physics \cite{Spreeuw1998,Paneru2020}.
Additionally, they help in understanding the boundaries between classical and quantum domains \cite{GellMann1993, Landsman2007}.

The exploration of these analogies traces its roots back to Schrödinger \cite{Masoliver2010}.
He introduced coherent states for the quantum harmonic oscillator as the state closest to classical theory \cite{Schrodinger1926}.
The theory of optical coherence \cite{Mandel1995} further reinforced the connection between classical and quantum optics.
It showed that the factorizable correlation functions of coherent states of the electromagnetic field behaved similarly to that of classical waves \cite{Glauber1963, Glauber1963b, Sudarshan1963}.
These classical statistical properties are fundamental in formulating the quantum theory of the electromagnetic field \cite{Mandel1995}. 
Interestingly, coherent states continue to play a crucial role in elucidating the transitions between classical and quantum realms \cite{Zurek1993, Berceanu1994,DeMartini2012}.

A proposal exists for a discrete optical analogue of Gilmore-Perelomov coherent states in coupled waveguide devices \cite{VillanuevaVergara2015}. 
The structures related to two-mode rotations, or $su(2)$ symmetry, provide an experimental platform to demonstrate coherent quantum transport \cite{PerezLeija2013, Espinosa2022} and the discrete fractional Fourier transform \cite{Tschernig2018}.
Recently, we showed optical analogues of coherent states using Laguerre-Gauss modes \cite{MoralesRodriguez2024}.
Here, we aim to extend the continuous variable analogy  by linking the superposition of classical optical Hermite-Gauss and Laguerre-Gauss modes to generalized quantum coherent states. 
This goal has significant implications for the development of robust photonic technologies \cite{He2022,Bliokh2023}. 
While quantum states of light hold immense potential for optical sensing, they are fragile when exposed to noise and decoherence \cite{Dowling2003}. 
In contrast, our classical optical analogues utilize different degrees of freedom within a single beam, enhancing their robustness for optical sensing. 
Similar to vector modes \cite{Hu2022}, the spatial modes we introduce demonstrate promising robustness, making them potentially suitable for sensing photosensitive materials.
In general, spatial light modes enable amplitude and phase transformations that produce classical optical analogues of nonclassical states. 
Consequently, our platform is also relevant for other technologies, including communication and information processing \cite{Lawrie2019}. 
At a more fundamental level, our approach bridges classical and quantum optics and highlights the potential of spatial light modes to simulate intricate quantum phenomena.

In this tutorial, we revisit the transverse modes of light and their physical characteristics within the paraxial approximation in Sec. \ref{sec:S2}.
We connect them to the eigenvalue problem for the two-dimensional isotropic quantum harmonic oscillator. 
In Section \ref{sec:S3}, we focus on the eigenvalue problem and its normal modes, constructing the Hermite- and Laguerre-Gauss modes.
We examine three fundamental symmetries for these spatial modes of light.
First, we study displacements that serve as the optical analogue of nonlinear coherent states in Sec. \ref{sec:S4}.
Second, we explore two-mode rotations that serve as the optical analogue of generalized spin coherent states in Sec. \ref{sec:S5}.
Third, we study two-mode squeezing that serves as the optical analogue of squeezed number states in Sec. \ref{sec:S6}.
For each case, we review the statistical properties of the corresponding generalized quantum coherent state using its corresponding Lie algebra.
We also construct their classical optical analogue using Hermite- and Laguerre-Gauss modes, discussing their irradiance and phase distributions. 
In Section \ref{sec:S7}, we provide the experimental methods to generate our optical analogues of generalized coherent states in the laboratory. 
We conclude in Sec. \ref{sec:S8}, discussing the relevance of our spatial light modes for optical sensing and metrology, as well as other optical technologies. 
We believe our work uncovers the potential of spatial light modes as an alternative platform for  exploring elusive physical effects in quantum multiparticle systems.
 
\section{Transverse spatial light modes } \label{sec:S2}

In isotropic homogeneous linear media, light is a transverse electromagnetic field \cite{VolkeSepulveda2006},
\begin{align}
    \mathbf{E}(\mathbf{r}) = c_{TE} \vec{\nabla} \times \left[ \mathbf{u} \psi(\mathbf{r}) \right] + c_{TM} \frac{1}{k} \vec{\nabla} \times \vec{\nabla} \times \left[ \mathbf{u} \psi(\mathbf{r}) \right]
\end{align}
that answers to Maxwell equations.
The complex parameters $c_{TE}$ and $c_{TM}$ relate to the transverse electric and magnetic components of the field, respectively.
The shorthand notation $\vec{\nabla}$ stands for the vector Laplacian operator.
Choosing the $z$-axis as the propagation direction,
\begin{align}
    \psi(\mathbf{r}) = f(\mathbf{r}_{\perp}) e^{i (k_{z} z - \omega t)},
\end{align}
reveals that the scalar spatial function,
\begin{align}
    \left( \nabla_{\perp}^{2} + k^2 \right) f(\mathbf{r}_{\perp})  = 0,
\end{align}
is invariant under propagation.
We use the total wave number $k = 2 \pi / \lambda$, its projection along the propagation direction $k_{z}$, and the field frequency $\omega = k v$ in terms of the speed of light in the medium $v$, and the shorthand notation $\nabla_{\perp}^{2}$ for the Laplacian in the plane transverse to propagation.
This is a regular Sturm-Liouville problem with Dirichlet boundary conditions \cite{Arfken2013}.
There exists an infinite number or real eigenvalues $k^{2}$, degenerate or not, that can be numbered in ascending order according to their value. 
For each eigenvalue, there exists a unique eigenfunction with a number of zeroes proportional to the label of the eigenvalues. 
These eigenfunctions form an orhtonormal basis for a Hilbert space that shows different underlying symmetries. 

This framework unveils the diverse degrees of freedom inherent to light modes. 
Polarization, akin to intrinsic angular momentum or spin, is determined by the vector $\mathbf{u}$. 
The longitudinal degree of freedom, aligned with the $z$-axis, provided by the plane wave $e^{i k_{z} z}$ conserves linear momentum.
The spatial transverse degree of freedom allows various representations. 
In Cartesian coordinates, it conserves total transverse linear momentum as ideal plane wave fields.
In polar coordinates, it conserves radial and orbital angular momentum as ideal Bessel fields \cite{Jauregui2005, Lloyd217}, leading to complex light structures like vortex fields.
More intricate representations conserving elliptic and parabolic momenta emerge in ideal Mathieu \cite{RodriguezLara2008, GutierrezVega2000} or Weber \cite{RodriguezLara2009} fields. 
Each photon in these ideal field modes carries a  quantized value of the respective constant of motion \cite{Jauregui2005,RodriguezLara2008,RodriguezLara2009}.
While these are idealized representations, more realistic fields are achievable through the superposition of these fundamental modes. 
This approach allows for the creation of tailored classical or quantum light fields suited for specific applications \cite{RodriguezLara2009b,PerezPascual2011, RosalesGuzman2018, Forbes2021}.

The paraxial approximation, $k_{z} \gg k_{\perp}$, is a practical alternative to construct realistic Gaussian beam light fields,
\begin{align}
    \mathbf{E}(\mathbf{r}) = \mathbf{u} \Psi(\mathbf{r}) e^{i (k z - \omega t)},
\end{align}
where polarization is decoupled from the spatial degree of freedom.
The standard slowly varying Gaussian scalar envelope solution,
\begin{align}
  \Psi(\mathbf{r}) &= \frac{w_{0}}{w(z)} e^{\frac{i k r^{2}}{2 R(z)}} e^{ -i (n+1) \varphi(z)} e^{ -\frac{r^{2}}{w^{2}(z)}} \psi(r_{\perp}),
\end{align}
relies on propagation-dependent beam parameters, 
\begin{align}
    \begin{aligned}
        w(z) =&~  w_{0} \sqrt{1 + \left(\frac{z}{z_{R}} \right)^{2}} , \\
        R(z) =&~ z \left[ 1 + \left(\frac{z_{R}}{z}\right)^{2}\right], \\
        \varphi(z) =&~ \tan^{-1}\frac{z}{z_{R}}, 
    \end{aligned}
\end{align}
defining the beam width, curvature radius, and Guoy phase, respectively.
These link to the constant parameters,
\begin{align}
    \begin{aligned}
        w_{0} =&~ \frac{\lambda}{\pi \theta}, \\
        z_{R} =&~ \frac{\pi}{\lambda} w_{0}^2,
    \end{aligned}
\end{align}
providing the waist radius and Rayleigh range, in that order, where the parameter $\theta$ is the beam divergence.
Here, the constant integer number $n$ plays an interesting role.
This Gaussian scalar wave satisfies the paraxial wave equation \cite{Lax1975},
\begin{align}
  \left(\nabla_{\perp}^{2} + 2 i k \partial_{z} \right) \Psi(\mathbf{r}) = 0,
\end{align}
where we use the shorthand notation  $\partial_{z}$ for the partial derivative along the propagation axis.
It is cumbersome but straightforward to show that the auxiliary transverse function fulfills a differential equation \cite{Nienhuis2004}, 
\begin{align}
    \frac{1}{2}\left(-\nabla_{\perp}^{2} + \tilde{r}_{\perp}^{2} \right) \psi\left( \tilde{\mathbf{r}}_{\perp} \right) &= (n+1) \psi \left( \tilde{\mathbf{r}}_{\perp} \right),
\end{align}
equivalent to the dimensionless eigenvalue problem for a dimensionless two-dimensional isotropic harmonic oscillator if we choose to scale the transverse coordinate vector as $\tilde{\mathbf{r}}_{\perp} = \sqrt{2} \, \mathbf{r}_{\perp} / w(z)$.
A finite version of this eigenvalue problem also appear in Parabolic gradient index fibers \cite{Collado2024}.
Hermite-Gaussian beams \cite{Siegman1986} are a set of solutions that shows linear momentum in both horizontal and vertical direction in the plane transverse to propagation.
Laguerre-Gaussian beams \cite{Allen1992, Nienhuis1993} are another set that shows orbital angular momentum and an intrinsic hyperbolic momentum \cite{Karimi2014,Plick2015}.
Again, this is a regular Sturm-Liouville problem with Dirichlet boundary conditions, providing a complete orthonormal basis for a Hilbert space with different underlying symmetries that we review in the following.

\section{The Sturm-Liouville problem} \label{sec:S3}

The dimensionless eigenvalue problem in operator form,
\begin{align}
     \frac{1}{2} \sum_{j=x,y} \left( \hat{p}_{j}^{2} +  \hat{q}_{j}^{2} \right) \psi_{n}(q_{x},q_{y}) = (n+1) \psi_{n}(q_{x},q_{y}),
\end{align}
with $n = 0, 1, 2,\ldots$, reduces to the two-dimensional isotropic harmonic oscillator eigenvalue problem \cite{Dennis2017}.
The action of the dimensionless canonical pair,
\begin{align}
   \begin{aligned}
      \hat{q}_{j} \,\psi_{n}(q_{x},q_{y}) &=~ q_{j} \, \psi_{n}(q_{x},q_{y}), \\
      \hat{p}_{j} \, \psi_{n}(q_{x},q_{y}) &=~ - i \partial_{q_{j}} \, \psi_{n}(q_{x},q_{y}),
   \end{aligned}
\end{align}
leads to the dimensionless canonical commutation relation for the Heisenberg-Weyl algebra \cite{Schmudgen2020},
\begin{align}
    \left[ \hat{q}_{j}, \hat{p}_{k} \right] = i \delta_{j,k} 
\end{align}
with the commutator operation $\left[ \hat{A}, \hat{B} \right] = \hat{A} \hat{B} - \hat{B} \hat{A} $ and Kronecker delta $\delta_{j,k}$.

From an abstract algebra point of view, it is helpful to define annihilation and creation operators \cite{mahan2013}, 
\begin{align}\label{LadderOperator}
    \begin{aligned}
            \hat{a}_{j} &=~ \frac{1}{\sqrt{2}} \left( \hat{q}_{j} + i \hat{p}_{j} \right), \\
            \hat{a}_{j}^{\dagger} &=~ \frac{1}{\sqrt{2}} \left( \hat{q}_{j} - i \hat{p}_{j} \right),
    \end{aligned}
\end{align}
that fulfill the corresponding commutation relation, 
\begin{align}
    \left[ \hat{a}_{j}, \hat{a}_{k}^{\dagger} \right] =  \delta_{j,k} 
\end{align}
with $j,k= x, y$.
In Cartesian configuration space, it is straightforward to identify the differential equation for zeroth eigenstate,
\begin{align}
    \frac{1}{2} \sum_{j=x,y} \left( - \partial_{q_{j}}^{2} +  q_{j}^{2} \right) \psi_{0}(q_{x},q_{y}) = \psi_{0}(q_{x},q_{y}),
\end{align}
that yields a Gaussian solution, 
\begin{align}
    \psi_{0}(q_{x},q_{y}) = \dfrac{1}{\sqrt{ \pi }}  ~e^{-\frac{1}{2} (q_{x}^2+q_{y}^2)}, 
\end{align}
that we will identify as the ground state.

At this point, we may define a generalized annihilation and creation operators,
\begin{align}
    \begin{aligned}
        \hat{a}(\theta, \phi) &=~ \hat{a}_{x} \cos \theta + \hat{a}_{y} e^{i \phi} \sin \theta, \\
        \hat{a}^{\dagger}(\theta, \phi) &=~ \hat{a}_{x}^{\dagger} \cos \theta + \hat{a}_{y}^{\dagger} e^{-i \phi} \sin \theta,
    \end{aligned}
\end{align}
following a construction similar to that of Jones polarization vectors \cite{Jones1941} with angle parameters $\theta \in \left[ 0, 2\pi \right)$ and $ \phi \in \left[ 0, \pi \right]$.
Thus, we may realize that the original eigenvalue problem, 
\begin{align}
    \begin{aligned}
        \frac{1}{2}\sum_{j=x,y} \left( \hat{p}_{j}^{2} +  \hat{q}_{j}^{2} \right) &=~ \frac{1}{2}\sum_{j=1,2} \left\{ \hat{a}_{j} ,  \hat{a}_{j}^{\dagger} \right\}, 
    \end{aligned}
\end{align}
with the anti-commutator operation $\{ \hat{A}, \hat{B} \} = \hat{A} \hat{B} + \hat{B} \hat{A} $, allows for infinitely many representations in terms of the commuting annihilation operators, 
\begin{align}
    \begin{aligned}
        \hat{a}_{1} \equiv&~ \hat{a}\left( \theta, \phi \right), \\
        \hat{a}_{2} \equiv&~ \hat{a}\left( \pi/2 +\theta, \phi \right),
    \end{aligned}
\end{align}
fulfilling the standard Heisenberg-Weyl algebra commutation relations \cite{Sakurai2011}, 
\begin{align}
    \left[ \hat{a}_{j}, \hat{a}_{k}^{\dagger} \right] =  \delta_{j,k} 
\end{align}
with $j,k = 1,2$, for the creation--annihilation operator pair.

These generalized creation and annihilation operators fulfill the commutation relations, 
\begin{align}
    \begin{aligned}
        \left[ \hat{a}_{j}^{\dagger} \hat{a}_{j} , \hat{a}_{j} \right] &=~ - \hat{a}_{j}, \\
        \left[ \hat{a}_{j}^{\dagger}\hat{a}_{j}, \hat{a}_{j}^{\dagger} \right] &=~ \hat{a}_{j}^{\dagger},
    \end{aligned}
\end{align}
with $j=1,2$, that allow us to realize that we can define the rest of the eigenstates, 
\begin{align}
    \begin{aligned}
        \vert n_{1}, n_{2} \rangle \equiv \frac{1}{\sqrt{n_{1}! n_{2}!} } \hat{a}_{1}^{\dagger n_{1} } \hat{a}_{2}^{\dagger n_{2} } \vert 0 \rangle,
    \end{aligned}
\end{align}
using a shorthand vector notation, where the action of the creation and annihilation operators,
\begin{align}
    \begin{aligned}
        \hat{a}_{j}  \vert n_{1},n_{2} \rangle &=~ \sqrt{n_{j}} \, \vert n_{1}-\delta_{j,1},n_{2}-\delta_{j,2} \rangle, \\
        \hat{a}_{j}^{\dagger}  \vert n_{1},n_{2} \rangle &=~ \sqrt{n_{j}+1} \, \vert n_{1}+\delta_{j,1},n_{2}+\delta_{j,2} \rangle, \\
        \hat{a}_{j}^{\dagger} \hat{a}_{j} \vert n_{1},n_{2} \rangle &=~ n_{j}  \, \vert n_{1},n_{2} \rangle,
    \end{aligned}
\end{align}
is derived from the commutation relation for the operators and the eigenvalue problem.
We favor a shorthand vector notation that we can use,
\begin{align}
    \langle q_{j}, q_{k} \vert n_{1}, n_{2} \rangle = \psi_{n_{1},n_{2}}(q_{j},q_{k}),
\end{align}  
for any given coordinate representation, $\{ j,k \} =  \{ x,y\}, \{ \rho, \varphi \}$, etc.
In the following, we will construct the normal modes for Cartesian and Polar configuration space.

\subsection{Hermite-Gauss modes}

In Cartesian configuration space, the generalized creation and annihilation operators yield eigenvectors, 
\begin{align}
    \vert n_{x}, n_{y} \rangle = \frac{1}{\sqrt{n_{x}! n_{y}!}}
    \hat{a}_{x}^{\dagger n_{x}}
    \hat{a}_{y}^{\dagger n_{y}} \vert 0 \rangle,
\end{align}
where it is possible to separate and write the action of the creation operator in differential form on the ground state  \cite{Sakurai2011, Zettili2009}, 
\begin{align}
    \left( q_{j} - \partial_{q_{j}} \right)^{n_{j}} e^{-\frac{1}{2} q_{j}^2} = e^{-\frac{1}{2} q_{j}^2}    \mathrm{H}_{n_{j}}(q_{j}) , 
\end{align}
with $j=x,y$, to recover the Hermite polynomials $\mathrm{H}_{n}(x)$  \cite{Strasburger2015,Arfken2013}.
Thus, the normal modes in Cartesian configuration space, 
\begin{align}
    \begin{aligned}
        \langle q_{x}, q_{y} \vert n_{x}, n_{y} \rangle &=~ \psi_{n_{x},n_{y}}(q_{x},q_{y}), \\
        &=~ \dfrac{1}{\sqrt{ \pi  2^{n_{x}+n_{y}}  n_{x}! n_{y}! \, }}  ~e^{-\frac{1}{2} (q_{x}^2+q_{y}^2)} \mathrm{H}_{n_{x}}(q_{x}) \mathrm{H}_{n_{y}}(q_{y} ) , 
    \end{aligned}
\end{align}
answer to the Sturm-Liouville problem with eigenvalue $n = n_{x} + n_{y}$.
The labels $n_{x}$ and $n_{y}$ are usually called the horizontal and vertical numbers.
The eigenvalues present $n+1$ degeneracy while these functions are orthonormal, 
\begin{align}
    \iint_{-\infty}^{\infty} dq_{x} dq_{y} \, \psi^{\ast}_{m_{x},m_{y}}(q_{x},q_{y}) \psi_{n_{x},n_{y}}(q_{x},q_{y}) = \delta_{m_{x},n_{x}} \delta_{m_{y},n_{y}},
\end{align}
and span a Hilbert space as expected.

\begin{figure}
\includegraphics[width = 0.75 \linewidth]{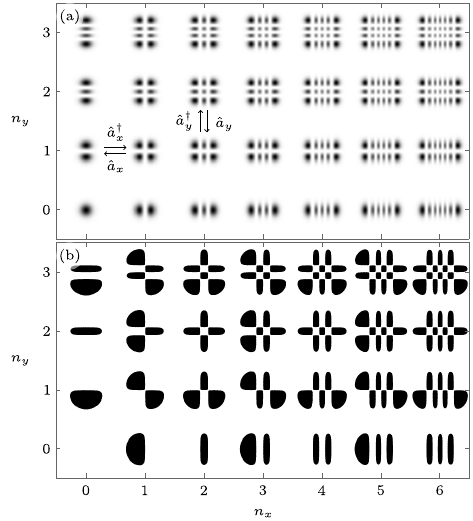}
\caption{(a) Irradiance $\vert \psi_{n_{x},n_{y}}(q_{x},q_{y}) \vert^{2} $ and (b) phase $\mathrm{arg} \left[ \psi_{n_{x},n_{y}}(q_{x},q_{y}) \right] $ distributions for some Hermite--Gauss modes.} \label{fig:FigHG}
\end{figure}

Figure \ref{fig:FigHG}(a) shows the irradiance distribution $\vert \psi_{n_{x},n_{y}}(q_{x},q_{y}) \vert^{2} $  where it can be seen that the horizontal or vertical label corresponds to the number of zeroes in that direction, while Fig. \ref{fig:FigHG}(b) shows its phase distribution $\mathrm{arg} \left[ \psi_{n_{x},n_{y}}(q_{x},q_{y}) \right] $. 
The action of the creation and annihilation operators on the Hermite-Gauss modes is shown in Fig. \ref{fig:FigHG}(a).

\subsection{Laguerre-Gauss modes}

In Polar configuration space, the generalized annihilation operators, 
\begin{align}
    \hat{a}_{\pm} &=~ \frac{1}{\sqrt{2}} \left( \hat{a}_{x} \mp i \hat{a}_{y} \right),
\end{align}
and corresponding creation operators, yield eigenvectors, 
\begin{align}
    \vert n_{+}, n_{-} \rangle = \frac{1}{\sqrt{n_{+}! n_{-}!} } \hat{a}_{+}^{\dagger n_{+} } \hat{a}_{-}^{\dagger n_{-} } \vert 0 \rangle,
\end{align}
that yield Laguerre-Gauss eigenfunctions, \cite{Schwinger2001}, 
\begin{align}
    \begin{aligned}
        \langle q_{\rho}, q_{\varphi} \vert n_{+}, n_{-} \rangle &=~ \psi_{n_{+},n_{-}}(q_{\rho},q_{\varphi}), \\
        &=~ \frac{1}{\sqrt{\pi }} (-1)^{p(n_{+},n_{-})} \sqrt{\frac{ p(n_{+},n_{-})!}{\left[p(n_{+},n_{-})+\vert n_{+} - n_{-} \vert\right]!}} \times\\
        &~ \times \, q_{\rho}^{\vert n_{+} - n_{-} \vert} e^{ - \frac{1}{2} q_{\rho}^{2}} \mathrm{L}_{p(n_{+},n_{-})}^{\vert n_{+} - n_{-} \vert} (q_{\rho}^{2}) e^{i (n_{+} - n_{-}) q_{\varphi}} , 
    \end{aligned}
\end{align}
that answer to the Sturm-Liouville problem with eigenvalue $n = n_{+} + n_{-}$ and $p(n_{+},n_{-}) = \left[ n_{+} + n_{-} - \vert n_{+} - n_{-} \vert \right]/2$.
The labels $n_{+}$ and $n_{-}$ are usually called the left and right circular numbers.
The eigenvalues present $n+1$ degeneracy while these functions are orthonormal, 
\begin{align}
    \int_{0}^{\infty} \int_{0}^{2 \pi} dq_{\rho} dq_{\varphi} \, q_{\rho} \psi^{\ast}_{m_{+},m_{-}}(q_{\rho},q_{\varphi}) \psi_{n_{+},n_{-}}(q_{\rho},q_{\varphi}) = \delta_{m_{+},n_{+}} \delta_{m_{-},n_{-}},
\end{align}
and span a Hilbert space as expected.

\begin{figure}
\includegraphics[width = 0.75 \linewidth]{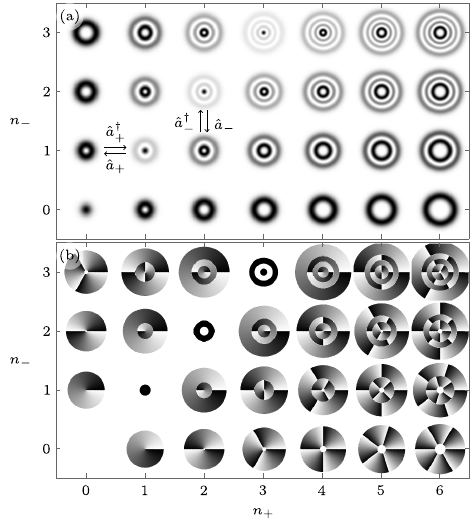}
\caption{(a) Irradiance $\vert \psi_{n_{+},n_{-}}(q_{\rho},q_{\vartheta}) \vert^{2} $ and (b) phase $\mathrm{arg}\left[ \psi_{n_{+},n_{-}}(q_{\rho},q_{\vartheta}) \right] $ distributions for some Laguerre--Gauss modes.} \label{fig:FigLG}
\end{figure}

Figure \ref{fig:FigLG}(a) shows the irradiance distribution $\vert \psi_{n_{+},n_{-}}(q_{\rho},q_{\vartheta}) \vert^{2} $ where it can be seen that the number of zeroes in the radial direction corresponds to the value of $p(n_{+},n_{-})$, while Fig. \ref{fig:FigLG}(b) shows its phase distribution $\mathrm{arg} \left[ \psi_{n_{+},n_{+}}(q_{\rho},q_{\vartheta}) \right]$, where it can be seen that the absolute value of the rest of the left and right circular numbers, $\vert n_{+} - n_{-} \vert$, corresponds to the number of times the phase winds from zero to $2\pi$. 
The action of the creation and annihilation operators on the Laguerre-Gauss modes is shown in Fig. \ref{fig:FigLG}(a).

\section{Displacements: Nonlinear coherent states} \label{sec:S4}

The Heisenberg-Weyl structure underlying the eigenvalue problem allows for the construction of unitary displacement operators \cite{Glauber1963},  
\begin{align}
    \begin{aligned}
        \hat{D}_{j}(q_{j0}, p_{j0}) = e^{ i \left( p_{j0} \hat{q}_{j} - q_{j0} \hat{p}_{j} \right) }
    \end{aligned}
\end{align}
with $j=1,2$, that receive its name from the fact that they  displace the generalized dimensionless canonical pair, 
\begin{align}
    \begin{aligned}
        \hat{D}_{j}^{\dagger}(q_{j0}, p_{j0}) \, \hat{q}_{k} \,  \hat{D}_{j}(q_{j0}, p_{j0}) =  \hat{q}_{k} + q_{k0}  \delta_{j,k}, \\
        \hat{D}_{j}^{\dagger}(q_{j0}, p_{j0}) \, \hat{p}_{k} \,  \hat{D}_{j}(q_{j0}, p_{j0}) =  \hat{p}_{k} + p_{k0}  \delta_{j,k},
    \end{aligned}
\end{align}
by a real constant $q_{j0}$ or $p_{j0}$. 
It is straightforward to use Zassenhaus formula for exponential maps of linear operators \cite{Casas2012},
\begin{align}
    \hat{D}_{j}(q_{j0},p_{j0}) = e^{-\frac{i}{2} q_{j0} p_{j0}} e^{i p_{j0} \hat{q}_{j} } e^{-i q_{j0} \hat{p}_{j} }   ,
\end{align}
and a McLaurin expansion of both exponential maps for the dimensional canonical pair to see, 
\begin{align}
    \begin{aligned}
        \hat{D}_{j}(q_{j0},p_{j0}) f(q_{1},q_{2}) &=~ e^{-\frac{i}{2} q_{j0} p_{j0}} e^{i p_{j0} \hat{q}_{j} } \times \\
        &~ \times f(q_{1} - q_{j0} \delta_{j,1}, q_{2} -  q_{j0} \delta_{j,2}),
    \end{aligned}
\end{align}
that this operation only translates and adds a constant momentum plane-wave term to any plausible state of the optical field.

The displacement operator in terms of the creation and annihilation operators \cite{Glauber1963}, 
\begin{align}
    \begin{aligned}
        \hat{D}_{j}(\alpha) = e^{ \alpha \hat{a}_{j}^{\dagger} - \alpha^{\ast} \hat{a}_{j}},
    \end{aligned}
\end{align}
provides complex displacements \cite{Cahill1969,Oliveira1990}, 
\begin{align}
    \begin{aligned}
        \hat{D}_{j}^{\dagger}(\alpha_{j}) \hat{a}_{k}  \hat{D}_{j}(\alpha_{j}) &= \hat{a}_{k} + \alpha_{k}  \delta_{j,k}, \\
        \hat{D}_{j}^{\dagger}(\alpha_{j}) \hat{a}_{k}^{\dagger}  \hat{D}_{j}(\alpha_{j}) &= \hat{a}^{\dagger}_{k} +\alpha_{k}^{\ast} \delta_{j,k},
    \end{aligned}
\end{align}
of the creation and annihilation operators where the complex constant relate to the real displacement constants, $ \mathrm{Re}(\alpha_{j}) = q_{j0}/ \sqrt{2}$ and $\mathrm{Im}(\alpha_{j}) = p_{j0}/\sqrt{2}$ that link the displacement operator in first and second quantization form. 

It is possible to displace any state.
In particular, displaced number states are known as nonlinear coherent states \cite{Cahill1969}, 
\begin{align}
    \begin{aligned}
        \vert \alpha_{1}, m_{1} , \alpha_{2}, m_{2} \rangle &=~    \hat{D}_{1}(\alpha_{1}) \hat{D}_{2}(\alpha_{2}) \vert m_{1}, m_{2} \rangle, \\
        &=~ \frac{(-1)^{m_{1}+m_{2}}}{\sqrt{m_{1}! m_{2}!}} e^{-\frac{1}{2} \left( \vert \alpha_{1} \vert^{2} + \vert \alpha_{2} \vert^{2}\right)} \alpha_{1}^{-m_{1}} \alpha_{2}^{-m_{2}} \times \\
        &~ \times \sum_{j,k=0}^{\infty} \frac{1}{\sqrt{j! k!}} \alpha_{1}^{j} \alpha_{2}^{k} \mathrm{U}\left(-m_{1},-m_{1}+j+1,\vert \alpha_{1}\vert^2 \right) \times \\
        &~ \times \mathrm{U}\left(-m_{2},-m_{2}+k+1,\vert \alpha_{2}\vert^2 \right) \vert j, k \rangle,
    \end{aligned}
\end{align}
in terms of Tricomi confluent hypergeometric function $U(a,b,c)$, such that in configuration space it takes the shape of the eigenstate displaced by their mean position and momentum,
\begin{align}
    \begin{aligned}
        \langle \hat{q}_{j} \rangle &=~ q_{j_{0}} = \sqrt{2} \vert \alpha_{j} \vert \cos \theta_{j}, \\
        \langle \hat{p}_{j} \rangle &=~ p_{j_{0}} = \sqrt{2} \vert \alpha_{j} \vert \sin \theta_{j}, 
    \end{aligned}
\end{align}
that are proportional to the complex displacement parameter $\alpha_{j} = \vert \alpha_{j} \vert e^{i \theta}$. 
The variances of the dimensionless canonical variables,
\begin{align}
    \begin{aligned}
        \sigma_{q_{j}}^{2} &=~ \langle \hat{q}_{j}^2 \rangle - \langle \hat{q}_{j} \rangle^{2} = m_{j} + \frac{1}{2}, \\
        \sigma_{p_{j}}^{2} &=~ \langle \hat{p}_{j}^2 \rangle - \langle \hat{p}_{j} \rangle^{2} = m_{j} + \frac{1}{2}, \\
    \end{aligned}
\end{align}
are proportional to the excitation number before the displacement and minimize Heisenberg uncertainty relation,
\begin{align}
    \sigma_{q_{j}}^{2} \sigma_{p_{j}}^{2} = \left( m_{j} + \frac{1}{2} \right)^{2},
\end{align}
for displaced vacuum, also known as Glauber-Sudarshan coherent state \cite{Glauber1963, Sudarshan1963}.

The nonlinear coherent states mean of the excitation number operators for the generalized coordinates \cite{Oliveira1990},
\begin{align}
    \langle \hat{n}_{j} \rangle = n_{j} + \vert \alpha_{j} \vert^{2},
\end{align}
is the initial excitation number before the displacement  plus the absolute value squared of the complex displacement parameter.
Their variance,
\begin{align}
    \sigma_{n_{j}}^2 = \left( 2 m_{j} + 1 \right) \vert \alpha_{j} \vert^{2},
\end{align}
is proportional to the initial excitation number before the displacement and to the absolute value of the coherent parameter.
The variance of the excitation number is equal to its mean value for the displaced vacuum state, that is, Glauber-Sudarshan coherent states.

\subsection{Hermite-Gauss modes}

The displaced Hermite-Gauss modes in Cartesian dimensionless configuration space take an intuitive form, 
\begin{align}
    \begin{aligned}
        \langle q_{x}, q_{y} \vert \alpha_{x}, m_{x}, \alpha_{y}, m_{y} \rangle  &= e^{-\frac{i}{2} \left( q_{x0} p_{x0} + q_{y0} p_{y0} \right)} e^{ i \left( p_{x0} \hat{q}_{x} + p_{y0} \hat{q}_{y} \right)}  \times \\
        &~ \times  \psi_{m_{x},m_{y}}(q_{x} - q_{x0}, q_{y} - q_{y0}) ,
    \end{aligned}
\end{align}
with dimensionless position and momentum displacements,
\begin{align}
    \begin{aligned}
        q_{j0} &= \sqrt{2} \, \vert \alpha_{j} \vert \cos \theta_{j}, \\
        p_{j0} &= \sqrt{2} \, \vert \alpha_{j} \vert \sin \theta_{j},
    \end{aligned}    
\end{align}
in terms of the complex displacement parameters $\alpha_{j} = \vert \alpha_{j} \vert e^{ i \theta_{j} }$ with $j=x,y$.
The real parts of the complex displacement parameters define the position of the irradiance centroid and their imaginary parts their phase distribution up to a constant phase proportional to the product of their real and imaginary components
The plane-wave-like phase terms in the displaced Hermite-Gauss modes play a fundamental for propagation of these states.

\begin{figure}
\includegraphics[width = 0.75 \linewidth]{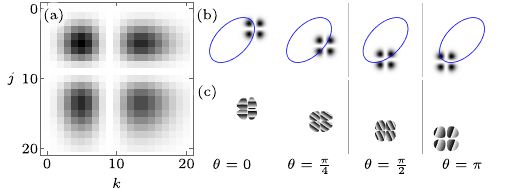}
\caption{(a) Probability $P_{j,k}(\alpha_{x},\alpha_{y} )$, (b) Irradiance $\vert \Psi_{\alpha_{x},\alpha_{y}} (q_{x}, q_{y}) \vert^{2}$, and (c) Phase $\mathrm{arg} [ \Psi_{\alpha_{x},\alpha_{y}} (q_{x}, q_{y}) ]$ distributions for nonlinear coherent states of a Hermite-Gauss mode with $n_{x}=n_{y}=1$ and complex coherent parameters $\alpha_{x} = \vert \alpha \vert e^{i \theta}$ and $\alpha_{y} = \vert \alpha \vert e^{i (\theta + \pi/3)}$ with amplitude $\vert \alpha \vert = 3$ and phase values $\theta = 0, \pi/4, \pi/2, \pi$.} \label{fig:FigHGDisp}
\end{figure}

The coordinate for their irradiance centroid in dimensionless Cartesian configuration space,
\begin{align}
    \sqrt{2} \, \vert \alpha_{x} \vert \, ( \cos \theta_{x} , \epsilon \cos \theta_{x} + \Delta \theta ),
\end{align}
where we define the ratio between the amplitudes of the complex displacement parameters $\epsilon = \vert \alpha_{y} \vert / \vert \alpha_{x} \vert$ and the complex displacement parameter phase difference $\Delta \theta = \theta_{y} - \theta_{x}$, defines an ellipse centered at the origin of dimensionless configuration space with semi-axes, 
\begin{align}
    \begin{aligned}
            a^{2} &=~ \frac{\epsilon \sin^{2} \vartheta - \cos^{2} \vartheta }{\sin^{4} \vartheta - \cos^{4} \vartheta }, \\
            b^{2} &=~ \frac{ \sin^{2} \vartheta - \epsilon \cos^{2} \vartheta }{\sin^{4} \vartheta - \cos^{4} \vartheta },
    \end{aligned}
\end{align}
in terms of inclination angle, 
\begin{align}
    \tan \vartheta = \frac{\sec \Delta \theta}{2 \epsilon} \left(  \epsilon^{2  } - 1 + \sqrt{1+ \epsilon^{2} (\epsilon^{2} + 2 \cos 2 \Delta \theta)} \right),
\end{align}
for the semi-axis $a$ with respect to the horizontal axis.
The complex displacement parameter amplitude ratio and phase difference defines the type of elliptical trajectory; for example, identical amplitudes and non-identical phases will locate the irradiance centroid on a ellipse, Fig. \ref{fig:FigHGDisp}(b) and  Fig. \ref{fig:FigHGDisp}(c), while identical phases do it on a line.  

\subsection{Laguerre-Gauss modes}

It is simpler to express the displaced Laguerre-Gauss modes in Cartesian dimensionless configuration,
\begin{align}
    \begin{aligned}
        \langle q_{\rho}, q_{\vartheta} \vert \alpha_{+}, m_{+}, \alpha_{-}, m_{-} \rangle  &= e^{-i \vert \alpha_{+} \alpha_{-}\vert \sin \left( \theta_{+} + \theta_{-}  \right)} e^{i \left( p_{x0} \hat{q}_{x} + p_{y0} \hat{q}_{y} \right)} \times \\
        &~ \times \psi_{m_{+},m_{-}} \left( q_{x} - q_{x0}, q_{y} - q_{y0} \right),
    \end{aligned}
\end{align}
with dimensionless position and momentum displacements, 
\begin{align}
    \begin{aligned}
        q_{x0} &=~ \vert \alpha_{+} \vert \cos \theta_{+} + \vert \alpha_{-} \vert \cos \theta_{-}, \\
        q_{y0} &=~ - \vert \alpha_{+} \vert \sin \theta_{+} + \vert \alpha_{-} \vert \sin \theta_{-}, \\
        p_{x0} &=~ \vert \alpha_{+} \vert \sin \theta_{+} + \vert \alpha_{-} \vert \sin \theta_{-}, \\
        p_{y0} &=~ \vert \alpha_{+} \vert \cos \theta_{+} - \vert \alpha_{-} \vert \cos \theta_{-}, 
    \end{aligned}
\end{align}
in terms of the complex displacement parameters $\alpha_{j} = \vert \alpha_{j} \vert e^{i \theta_{j}}$ with $j=\pm$.
The real parts of the addition and subtraction of the complex displacement parameters define the position of the irradiance centroid and their imaginary parts their phase distribution up to a constant phase proportional to the product of their amplitudes times the cosine of their added phases. 
The plane-wave-like terms in the displaced Laguerre-Gauss modes proportional to the imaginary parts of the addition and subtraction of the complex displacement parameters play a fundamental role for propagation of these states.

\begin{figure}
\includegraphics[width = 0.75 \linewidth]{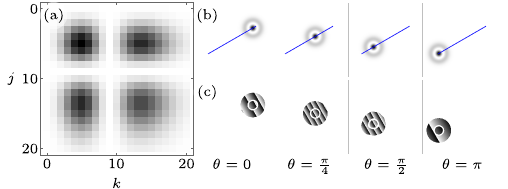}
\caption{(a) Probability $P_{j,k}(\alpha_{+},\alpha_{-} )$, (b) Irradiance $\vert \Psi_{\alpha_{+},\alpha_{-}} (q_{x}, q_{y}) \vert^{2}$, and (c) Phase $\mathrm{arg} \left[ \Psi_{\alpha_{+},\alpha_{-}} (q_{x}, q_{y}) \right]$ distributions for nonlinear coherent states of a Laguerre-Gauss mode with $n_{+}=n_{-}=1$ and complex coherent parameters $\alpha_{+} = \vert \alpha \vert e^{i \theta}$ and $\alpha_{-} = \vert \alpha \vert e^{i (\theta + \pi/3)}$ with amplitude $\vert \alpha \vert = 3$ and phase values $\theta = 0, \pi/4, \pi/2, \pi$.} \label{fig:FigLGDisp}
\end{figure}

The coordinate for their irradiance centroid in dimensionless Cartesian configuration space,
\begin{align}
     ( \vert \alpha_{+} \vert \cos \theta_{+} + \vert \alpha_{-} \vert \cos \theta_{-} , -\vert \alpha_{+} \vert \sin \theta_{+} + \vert \alpha_{-} \vert \sin \theta_{-} ),
\end{align}
defines an ellipse centered at the origin of dimensionless configuration space with semi-axes, 
\begin{align}
    \begin{aligned}
        a &=~ \vert \alpha_{+} \vert^{2} + \vert \alpha_{-} \vert^{2} , \\
        b &=~ \vert \alpha_{+} \vert^{2} - \vert \alpha_{-} \vert^{2}, 
    \end{aligned}
\end{align}
with the inclination angle, 
\begin{align}
    \tan \vartheta = - \cot \frac{\Delta}{2},
\end{align}
for the semi-axes $a$ with respect to the horizontal axis.
It is straightforward to realize that the irradiance centroid lies on a line for complex displacement parameters with the same amplitude, Fig. \ref{fig:FigLGDisp}, on a circle for one of them is zero, and an ellipse for any other parameter ratio.

\section{Rotations: Generalized spin coherent states} \label{sec:S5}

Let us define a two-boson representation of dimensionless angular momentum \cite{Schwinger1952}, 
\begin{align}
    \begin{aligned}
        \hat{J}_{1} &=\frac{1}{2} \left( \hat{a}_{1} \hat{a}_{2}^{\dagger} + \hat{a}_{1}^{\dagger} \hat{a}_{2} \right), \\
        \hat{J}_{2} &= \frac{i}{2} \left(  \hat{a}_{1} \hat{a}_{2}^{\dagger} - \hat{a}_{1}^{\dagger} \hat{a}_{2}  \right), \\
        \hat{J}_{3} &=  \frac {1}{2} \left(\hat {a}_{1}^{\dagger}  \hat{a}_{1} - \hat{a}_{2}^{\dagger}  \hat{a}_{2}\right),
    \end{aligned}
\end{align}
yielding the special unitary $su(2)$ Lie algebra,
\begin{align}
    \begin{aligned}
        \left[ \hat{J}_{i}, \hat{J}_{j} \right]&=i \epsilon_{ijk} \hat{J}_{k},
    \end{aligned}
\end{align}
with the Levi-Civita symbol $\epsilon_{ijk}$, accepting an alternate representation,
\begin{align}
    \begin{aligned}
        \left[\hat{J}_{3}, \hat{J}_{\pm}\right] &= \pm \hat{J}_{\pm},\\
        \left[\hat{J}_{+},\hat{J}_{-}\right] &= 2 \hat{J}_{3},
    \end{aligned}
\end{align}
in terms of the ladder operators \cite{Sakurai2011}, 
\begin{align}
    \begin{aligned}
        \hat{J}_{+} &= \hat{J}_{1} + i \hat{J}_{2} = \hat{a}_{1}^{\dagger} \hat{a}_{2}, \\
        \hat{J}_{-} &= \hat{J}_{1} - i \hat{J}_{2} = \hat{a}_{1} \hat{a}_{2}^{\dagger},
    \end{aligned}
\end{align}
It is important to stress that the algebra has a Casimir invariant,
\begin{align}
    \begin{aligned}
        \hat{J}^{2} &= \hat{J}_{1}^{2} + \hat{J}_{1}^{2} + \hat{J}_{3}^{2}, \\
        &= \hat{J}_{3}^{2} + \frac{1}{2} \left( \hat{J}_{+} \hat{J}_{-} + \hat{J}_{-} \hat{J}_{+} \right), \\
        &=  \left( \frac{\hat{a}_{1}^{\dagger} \hat{a}_{1} + \hat{a}_{2}^{\dagger} \hat{a}_{2}}{2} \right) \left( \frac{\hat{a}_{1}^{\dagger} \hat{a}_{1} + \hat{a}_{2}^{\dagger} \hat{a}_{2}}{2}  + 1 \right),
    \end{aligned}
\end{align}
that commutes, 
\begin{align}
    \left[ \hat{J}^{2}, \hat{J}_{k} \right] = 0
\end{align}
with all other elements of the algebra in any representation, $k = 1,2,3, \pm$.

It is possible to use a two-label basis that simultaneously diagonalize the Casimir $\hat{J}^{2}$ operator and one of the other operators, for example the so-called Dicke basis \cite{Ban1993, Vourdas1990, Arecchi1972},
\begin{align}
    \begin{aligned}
        \hat{J}^{2} \vert j; m \rangle &= j (j+1) \vert j; m \rangle, \\
        \hat{J}_{3} \vert j; m \rangle &= m \vert j; m \rangle, \\
        \hat{J}_{\pm} \vert j; m \rangle &= \sqrt{ (j \mp m) ( j \pm m +1)} \vert j; m \pm 1\rangle,
    \end{aligned}
\end{align}
with ladder operators $\hat{J}_{\pm}$.
The first label, the Bargmann parameter $j = 0, 1/2, 1, 3/2, \ldots$, is related to the Casimir operator and, in consequence, the total excitation number on this basis $\langle \hat{N} \rangle = \langle \hat{a}_{1}^{\dagger} \hat{a}_{1} + \hat{a}_{2}^{\dagger} \hat{a}_{2} \rangle = 2 j$, and the second label to the excitation number $ m = -j, j+1, \ldots, j-1, j$, such that there are $2j + 1$ excitation states in each Bargmann subspace.
Thus, the connection between Dicke states and excitation number states,
\begin{align}
    \vert j; m \rangle = \vert n_{1} = j + m, n_{2} = j - m \rangle,
\end{align}
is straightforward to make.

The $su(2)$ Lie algebra structure underlying the eigenvalue problem allows for the construction of a unitary rotation \cite{Arecchi1972}, 
\begin{align}
    \begin{aligned}
        \hat{R}(\alpha) &= e^{\left( \alpha \hat{J}_{+} - \alpha^{\ast}  \hat{J}_{-} \right)}, \\
        &= e^{ i \theta \hat{J}_{3}} e^{ i 2 \vert \alpha \vert \hat{J}_{2}} e^{-i \theta \hat{J}_{3}},
    \end{aligned}
\end{align}
with complex coherent parameter $\alpha = \vert \alpha \vert e^{i \theta}$ that produces two rotations on the generalized dimensionless canonical pairs, 
\begin{align}
    \begin{aligned}
        \hat{R}^{\dagger}(\alpha) \, \hat{q}_{1} \, \hat{R}(\alpha) &= \hat{q}_{1} \cos \vert \alpha \vert + \hat{q}_{2} \sin \vert \alpha \vert \cos \theta - \hat{p}_{2} \sin \vert \alpha \vert \sin \theta  , \\
        \hat{R}^{\dagger}(\alpha) \, \hat{q}_{2} \, \hat{R}(\alpha) &= \hat{q}_{2} \cos \vert \alpha \vert - \hat{q}_{1} \sin \vert \alpha \vert \cos \theta - \hat{p}_{1} \sin \vert \alpha \vert \sin \theta , \\
        \hat{R}^{\dagger}(\alpha) \, \hat{p}_{1} \, \hat{R}(\alpha) &= \hat{p}_{1} \cos \vert \alpha \vert \cos \theta + \hat{p}_{2} \sin \vert \alpha \vert - \hat{q}_{1} \cos \vert \alpha \vert \sin \theta , \\
        \hat{R}^{\dagger}(\alpha) \, \hat{p}_{2} \, \hat{R}(\alpha) &=  \hat{p}_{2} \cos \vert \alpha \vert \cos \theta - \hat{p}_{1} \sin \vert \alpha \vert + \hat{q}_{2} \cos \vert \alpha \vert \sin \theta, 
    \end{aligned}
\end{align}
proportional to the amplitude and phase of the coherent parameter.
Sadly, it is too cumbersome to calculate the action of the unitary $su(2)$ rotation on a function of dimensionless configuration space as we did for the unitary displacement for a complex coherent parameter. 
However, for a real coherent parameter, it is straightforward to see that the action in configuration space is a clockwise rotation by an angle equal to the real coherent parameter.

The rotation operator in terms of the creation and annihilation operators,  
\begin{align}
    \hat{R}(\alpha) &= e^{\left( \alpha \hat{a}_{1}^{\dagger} \hat{a}_{2} - \alpha^{\ast}  \hat{a}_{1} \hat{a}_{2}^{\dagger} \right)}, 
\end{align}
provides a rotation with a complex phase, 
\begin{align}
    \begin{aligned}
        \hat{R}^{\dagger}(\alpha) \hat{a}_{1}  \hat{R}(\alpha) &= \hat{a}_{1} \cos \vert \alpha \vert + \hat{a}_{2} e^{i \theta} \sin \vert \alpha \vert , \\
        \hat{R}^{\dagger}(\alpha) \hat{a}_{2}  \hat{R}(\alpha) &= \hat{a}_{2} \cos \vert \alpha \vert - \hat{a}_{1} e^{-i \theta} \sin \vert \alpha \vert, \\
        \hat{R}^{\dagger}(\alpha) \hat{a}_{1}^{\dagger}  \hat{R}(\alpha) &=  \hat{a}_{1}^{\dagger} \cos \vert \alpha \vert + \hat{a}_{2}^{\dagger} e^{- i \theta} \sin \vert \alpha \vert  ,\\
        \hat{R}^{\dagger}(\alpha) \hat{a}_{2}^{\dagger}  \hat{R}(\alpha) &= \hat{a}_{2}^{\dagger} \cos \vert \alpha \vert - \hat{a}_{1}^{\dagger} e^{i \theta} \sin \vert \alpha \vert,
    \end{aligned}
\end{align}
of the creation and annihilation operators where the amplitude of the complex parameter provides the rotation of the annihilation and creation operators. 

It is possible to rotate any state in the Dicke basis using  formulas for the exponential map of $su(2)$ operators in normal ordering \cite{Ban1993}. 
Rotated Dicke states are known as generalized spin coherent states \cite{VillanuevaVergara2015}, 
\begin{align}
    \begin{aligned}
        \vert j; \alpha, m \rangle &= \hat{R}(\alpha) \vert j; m \rangle, \\
        &= \sqrt{\left( \begin{array}{c} 2j \\ j-m \end{array} \right)} \, \sec^{-2 j} \vert \alpha \vert  \left( e^{ i \theta} \tan \vert \alpha \vert \right)^{j} \times \\
        &\times \sum_{k=0}^{2j} (-1)^{k} \sqrt{ \left( \begin{array}{c} 2j \\ k \end{array} \right)  } \,  e^{ - i \theta (k+m)} \tan^{(k-m)} \vert \alpha \vert   \,  \times \\
        & \times  \,_{2}\mathrm{F}_{1} \left(-k,-j+m; -2j; \csc^{2} \vert \alpha \vert \right) \, \vert j; j-k \rangle, 
    \end{aligned}
\end{align}
where we used the ordinary hypergeometric function $ \,_{2}\mathrm{F}_{1} \left(a,b;c; z\right)$ and binomial coefficients. 
This expression may be reduced to one in terms of Kravchuk polynomials \cite{VillanuevaVergara2015}.
These generalized spin coherent states in configuration space take a shape dependent on the representation used. 
However, it is possible to calculate the mean position and momentum for these generalized spin coherent states, 
\begin{align}
    \langle \hat{q}_{1} \rangle = 
    \langle \hat{q}_{2} \rangle = 
    \langle \hat{p}_{1} \rangle = 
    \langle \hat{p}_{2} \rangle = 0,
\end{align}
that is, their irradiance and phase centroids will stay at the origin, with identical variances for each dimensionless canonical pair, 
\begin{align}
    \begin{aligned}
        \sigma^{2}_{{q}_{1}} &= \sigma^{2}_{{p}_{1}}  = j + m \cos{2 \vert \alpha \vert} + \frac{1}{2},\\
        \sigma^{2}_{{q}_{2}} &= \sigma^{2}_{{p}_{2}}  = j - m \cos{2 \vert \alpha \vert} + \frac{1}{2},
    \end{aligned}
\end{align}
that produce Heisenberg uncertainty relations for each canonical pair, 
\begin{align}
    \begin{aligned}
        \sigma^{2}_{q_{1}} \sigma^{2}_{p_{1}} &= \left(j + m \cos{2 \vert \alpha \vert} + \frac{1}{2}\right)^{2}, \\
        \sigma^{2}_{q_{2}} \sigma^{2}_{p_{2}} &= \left(j - m \cos{2 \vert \alpha \vert} + \frac{1}{2}\right)^{2},
    \end{aligned}
\end{align}
that the generalized spin coherent state with $j=m=0$ minimize as expected as it is a Gaussian distribution.
For Bloch coherent states \cite{Arecchi1972, Radcliffe1971, Lieb2014}, that is $m = \pm j$, the uncertainty relations are minimized $\Delta q_{j} \Delta p_{j} = 1/2$ or maximized $\Delta q_{j} \Delta p_{j} = 2j + 1/2$ for coherent parameter amplitudes equal to half integer values of $\pi$, $\alpha = n \pi /2$ with $n=0,1,2, \ldots$
A minimum or maximum for any given value of the parameters $j$ and $m$ occurs at these values of the coherent parameter amplitude.

The generalized spin coherent states mean expected value for the excitation number operators in the generalized dimensionless coordinates, 
\begin{align}
    \begin{aligned}
        \langle \hat{n}_{1} \rangle &= j + m \cos{2 \vert \alpha \vert},\\
        \langle \hat{n}_{2} \rangle &=  j - m \cos{2 \vert \alpha \vert} ,\\
    \end{aligned}
\end{align}
provides us with a constant excitation number equal to two times the Bargmann parameter as expected, with variances, 

\begin{align}
    \begin{aligned}
        \sigma_{n_{1}}^{2}  &= \sigma_{n_{2}}^{2} = \frac{1}{2} (j + j^{2} - m^{2}) \sin^{2}{2 \vert \alpha \vert},
    \end{aligned}
\end{align}

that are equal to their expectation values only for the Bloch coherent states with $j=m=0$.
They take zero value for Bloch coherent states, $m=\pm j$, for coherent parameter amplitude equal to half integer values of $\pi$, $\alpha = n \pi /2$ with $n=0,1,2, \ldots$. 

The generalized spin coherent state mean expectation value for the elements of the $su(2)$ algebra, 
\begin{align}
    \begin{aligned}
        \langle \hat{J}_{1} \rangle &= -m \sin{2 \vert \alpha \vert} \cos{\theta}, \\
        \langle \hat{J}_{2} \rangle &= m \sin{2 \vert \alpha \vert} \sin{\theta}, \\
        \langle \hat{J}_{3} \rangle &= m \cos{2 \vert \alpha \vert}, 
    \end{aligned}
\end{align}
that may be seen as describing the coordinates of a point on a sphere of radius $m$. 
The variances for the expected values of these operators, 
\begin{align}
    \begin{aligned}
        \sigma^{2}_{J_{1}} &= \frac{1}{2}\left( j^{2} + j - m^{2} \right) \left( \cos^{2}{2 \vert \alpha \vert} \cos^{2}{\theta} + \sin^{2}{\theta}\right), \\
        \sigma^{2}_{J_{2}} &= \frac{1}{2}\left( j^{2} + j - m^{2} \right) \left( \cos^{2}{2 \vert \alpha \vert} \sin^{2}{\theta} + \cos^{2}{\theta}\right), \\
        \sigma^{2}_{J_{3}} &= \frac{1}{2}\left( j^{2} + j - m^{2} \right) \sin^{2}{2 \vert \alpha \vert}, 
    \end{aligned}
\end{align}
that are equal to their expectation values only for $j=m=0$.
For Bloch coherent states, $m=\pm j$,
the uncertainty relations are minimized $\Delta J_{1} \Delta J_{2} = j/2 $ for coherent parameter amplitudes equal to half integer values of $\pi$, $\alpha = n \pi /2$ with $n=0,1,2, \ldots$.

\subsection{Hermite-Gauss modes}

\begin{figure}
\includegraphics[width = 0.75 \linewidth]{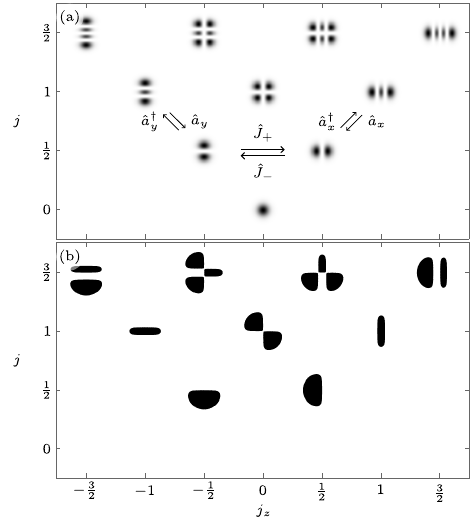}
\caption{(a) Irradiance and (b) phase distributions for Hermite--Gauss modes in two-mode $su(2)$ subspaces with constant $\langle \hat{J}^{2} \rangle = j (j+1)$.
Each line is a finite subspace of dimension $2j+1$ with Bargmann number $j=0,1/2,1,3/2, \ldots$} \label{fig:FigHGSU2}
\end{figure}

The Dicke modes in Cartesian configuration space, 
\begin{align}
    \begin{aligned}
        \langle q_{x}, q_{y} \vert j; m \rangle &=~ \psi_{j+m,j-m}(q_{x},q_{y}), \\
        &=~ \dfrac{e^{-\frac{1}{2} (q_{x}^2+q_{y}^2)}}{\sqrt{ \pi  2^{2j}  (j+m)! (j-m)! \, }}  ~ \mathrm{H}_{j+m}(q_{x}) \mathrm{H}_{j-m}(q_{y} ) , 
    \end{aligned}
\end{align}
answer to the Sturm-Liouville problem with eigenvalue $n = 2j$ that are $2j+1$ degenerate.
Thus, we may partition the whole Hilbert space into subspaces of constant total excitation number, that is, constant mean expectation value of the Casimir invariant, labelled by the Bargmann parameter $j$.
Figures \ref{fig:FigHGSU2}(a) and \ref{fig:FigHGSU2}(b) show the irradiance distribution $\vert \psi_{j+m,j-m}(q_{x},q_{y}) \vert^{2}$ and its phase distribution $\mathrm{arg} \left[ \psi_{j+m,j-m}(q_{x},q_{y}) \right]$, in that order.
Each line is a subspace of constant Bargmann parameter $j$ with $2j +1$ elements. 
Under this underlying symmetry, the action of the $su(2)$ ladder operators moves us left or right in each line while the action of the creation and annihilation operators of the Heisenberg-Weyl Lie algebra moves us in diagonal manner in the state ladder, \ref{fig:FigHGSU2}(a).

\begin{figure}
\includegraphics[width = 0.75 \linewidth]{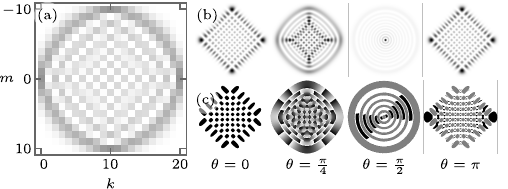}
\caption{(a) Probability $P_{k}(\alpha, m)$, (b) Irradiance $\vert \Psi_{\alpha, m} (q_{x}, q_{y}) \vert^{2}$, and (c) Phase $\mathrm{arg} \left[ \Psi_{\alpha, m} (q_{x}, q_{y}) \right]$ distributions for generalized spin coherent states of a Hermite-Gauss mode with $j=10$, $m=0$,  and complex coherent parameters $\alpha = \vert \alpha \vert e^{i \theta}$ with amplitude $\vert \alpha \vert = \pi/4$ and phase values $\theta = 0, \pi/4, \pi/2, \pi$.} \label{fig:FigHGRot}
\end{figure}

The generalized spin coherent state modes in Cartesian dimensionless configuration space take a non-intuitive form, 
\begin{align}
    \begin{aligned}
        \langle q_{x}, q_{y} \vert j; \alpha, m \rangle &= \sqrt{\left( \begin{array}{c} 2j \\ j-m \end{array} \right)} \, \sec^{-2 j} \vert \alpha \vert  \left( e^{ i \theta} \tan \vert \alpha \vert \right)^{j} \times \\
        &\times \sum_{k=0}^{2j} (-1)^{k} \sqrt{ \left( \begin{array}{c} 2j \\ k \end{array} \right)  } \,  e^{ - i \theta (k+m)} \tan^{(k-m)} \vert \alpha \vert   \,  \times \\
        & \times  \,_{2}\mathrm{F}_{1} \left(-k,-j+m; -2j; \csc^{2} \vert \alpha \vert \right) \,  \times \\
        & \times \dfrac{e^{-\frac{1}{2} (q_{x}^2+q_{y}^2)}}{\sqrt{ \pi  2^{2j}  (2j-k)! k! \, }}  ~ \mathrm{H}_{2j-k}(q_{x}) \mathrm{H}_{k}(q_{y} ) , 
    \end{aligned}
\end{align}
with their irradiance centroid at the origin of configuration space. 
Figure \ref{fig:FigHGRot}(a) shows the probability distribution for different initial Dicke states with excitation number parameter $m$.
Figure \ref{fig:FigHGRot}(b) and Fig. \ref{fig:FigHGRot}(c) show the irradiance distribution for a sample of these states.
As mentioned before, a fixed real coherent parameter, $\theta = 0, \pi$, provides a clock-wise rotation of the original Dicke state, first and last column of Fig. \ref{fig:FigHGRot}(b) and Fig. \ref{fig:FigHGRot}(c).

\subsection{Laguerre-Gauss modes}

\begin{figure}
\includegraphics[width = 0.75 \linewidth]{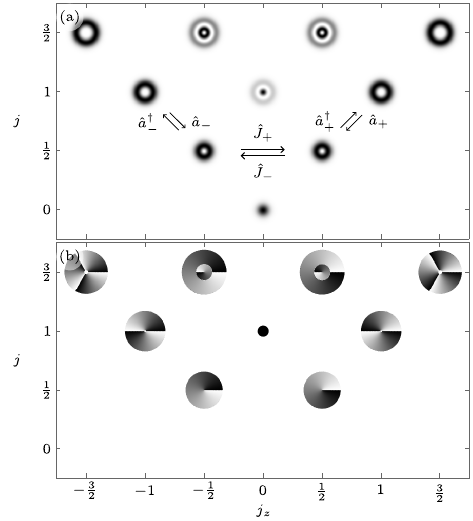}
\caption{(a) Irradiance and (b) phase distributions for Laguerre--Gauss modes in two-mode $su(2)$ subspaces with constant $\langle \hat{J}^{2} \rangle = j (j+1)$.
Each line is a finite subspace of dimension $2j+1$ with Bargmann number $j=0,1/2,1,3/2, \ldots$} \label{fig:FigLGSU2}
\end{figure}

The Dicke modes in Polar configuration space, 
\begin{align}
    \begin{aligned}
        \langle q_{\rho}, q_{\varphi} \vert j;m \rangle &=~ \psi_{j+m,j-m}(q_{\rho},q_{\varphi}), \\
        &=~ \frac{1}{\sqrt{\pi }} (-1)^{j - \vert m \vert} \sqrt{\frac{ \left( j - \vert m \vert \right)!}{\left( j + \vert m \vert  \right)!}} \times\\
        &~ \times \, q_{\rho}^{\vert 2 m \vert} e^{ - \frac{1}{2} q_{\rho}^{2}} \mathrm{L}_{j - \vert m \vert}^{\vert 2m \vert} (q_{\rho}^{2}) e^{i 2 m q_{\varphi}} , 
    \end{aligned}
\end{align}
answer to the Sturm-Liouville problem with eigenvalue $n = n_{+}+n_{-}= 2j$ that are $2j+1$ degenerate, where $n_{\pm} = \mathrm{min}(n_{+},n_{-}) + (\vert \ell \vert \pm \ell)/2$.
Thus, we may partition the whole Hilbert space into subspaces of constant total excitation number, that is, constant mean expectation value of the Casimir invariant, labelled by the Bargmann parameter $j$.
Figures \ref{fig:FigLGSU2}(a) and \ref{fig:FigLGSU2}(b) show the irradiance distribution $\vert \psi_{j+m,j-m}(q_{x},q_{y}) \vert^{2}$ and its phase distribution $\mathrm{arg} \left[ \psi_{j+m,j-m}(q_{x},q_{y}) \right]$, in that order.
Each line is a subspace of constant Bargmann parameter $j$ with $2j +1$ elements. 
Under this underlying symmetry, the action of the $su(2)$ ladder operators moves us left or right in each line while the action of the creation and annihilation operators of the Heisenberg-Weyl Lie algebra moves us in diagonal manner in the state ladder, \ref{fig:FigLGSU2}(a).

\begin{figure}
\includegraphics[width = 0.75 \linewidth]{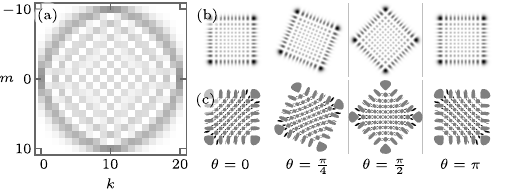}
\caption{(a) Probability $P_{k}(\alpha, m)$, (b) Irradiance $\vert \Psi_{\alpha, m} (q_{x}, q_{y}) \vert^{2}$, and (c) Phase $\mathrm{arg} \left[ \Psi_{\alpha, m} (q_{x}, q_{y}) \right]$ distributions for generalized spin coherent states of a Laguerre-Gauss mode with $j=10$, $m=0$,  and complex coherent parameters $\alpha = \vert \alpha \vert e^{i \theta}$ with amplitude $\vert \alpha \vert = \pi/4$ and phase values $\theta = 0, \pi/4, \pi/2, \pi$.} \label{fig:FigLGRot}
\end{figure}

The generalized spin coherent state modes in Polar dimensionless configuration space take a non-intuitive form, 
\begin{align}
    \begin{aligned}
        \langle q_{\rho}, q_{\varphi} \vert j; \alpha, m \rangle &= \sqrt{\left( \begin{array}{c} 2j \\ j-m \end{array} \right)} \, \sec^{-2 j} \vert \alpha \vert  \left( e^{ i \theta} \tan \vert \alpha \vert \right)^{j} \times \\
        &\times \sum_{k=0}^{2j} (-1)^{k} \sqrt{ \left( \begin{array}{c} 2j \\ k \end{array} \right)  } \,  e^{ - i \theta (k+m)} \tan^{(k-m)} \vert \alpha \vert   \,  \times \\
        & \times  \,_{2}\mathrm{F}_{1} \left(-k,-j+m; -2j; \csc^{2} \vert \alpha \vert \right) \,  \times \\
        & \times \frac{1}{\sqrt{\pi }} (-1)^{j - \vert j-k \vert} \sqrt{\frac{ \left( j - \vert j-k \vert \right)!}{\left( j + \vert j-k \vert  \right)!}} \times\\
        &~ \times \, q_{\rho}^{\vert 2 (j-k)) \vert} e^{ - \frac{1}{2} q_{\rho}^{2}} \mathrm{L}_{j - \vert j-k \vert}^{\vert 2(j-k) \vert} (q_{\rho}^{2}) e^{i 2 (j-k) q_{\varphi}} ,
    \end{aligned}
\end{align}
with their irradiance centroid at the origin of configuration space. 
Figure \ref{fig:FigLGRot}(a) shows the probability distribution for different initial Dicke states with excitation number parameter $m$.
Figure \ref{fig:FigLGRot}(b) and Fig. \ref{fig:FigLGRot}(c) show the irradiance distribution for a sample of these states.
Here, half the complex coherent parameter phase determines the angle for a clockwise rotation of the irradiance and phase distributions. 
Complex coherent parameter amplitude values $\vert \alpha \vert = \pi/4, 3 \pi / 4, 5 \pi/4 , 7 \pi /4$ produce Hermite-Gauss modes; for example, an amplitude $\vert \alpha \vert = \pi/4$, yields a mode with $n_{x} = j + m$ and $n_{y} = j - m$.
Complex coherent parameter amplitude values $\vert \alpha \vert = \pi/2, 3 \pi / 2$ produces circular structures with elliptic structures in between the former and these.

\section{Squeezing: Squeezed number states} \label{sec:S6}

Let us define a two-boson representation related to the symmetires of 1+1D psuedo-Euclidean space, 
\begin{align}
    \begin{aligned}
        \hat{K}_{1} &= \frac{1}{2} \left(\hat{a}_{1}^{\dagger} \hat{a}_{2}^{\dagger} + \hat{a}_{1} \hat{a}_{2}\right), \\
        \hat{K}_{2} &= -\frac{i}{2} \left(\hat{a}_{1}^{\dagger} \hat{a}_{2}^{\dagger} - \hat{a}_{1} \hat{a}_{2} \right), \\
        \hat{K}_{3} &= \frac{1}{2}\left(\hat{a}_{1}^{\dagger} \hat{a}_{1} + \hat{a}_{2}^{\dagger} \hat{a}_{2} +1 \right),
    \end{aligned}
\end{align}
yielding the special unitary $su(1,1)$ Lie algebra \cite{Wodkiewicz1985,Brif1996,Gerry1988},
\begin{align}
    \left[\hat{K}_{1}, \hat{K}_{2}\right]= -i \hat{K}_{3} , \quad
    \left[\hat{K}_{2}, \hat{K}_{3}\right]= i \hat{K}_{1}, \quad
    \left[\hat{K}_{3}, \hat{K}_{1}\right]= i \hat{K}_{2}, 
\end{align}
accepting an alternate representation,
\begin{align}
    \begin{aligned}
        \left[\hat{K}_{3}, \hat{K}_{\pm}\right] &= \pm \hat{K}_{\pm},\\
        \left[\hat{K}_{+},\hat{K}_{-}\right] &= -2 \hat{K}_{3},
    \end{aligned}
\end{align}
in terms of the ladder operators \cite{Shaterzadeh2008, Sakurai2011}, 
\begin{align}
    \begin{aligned}
        \hat{K}_{+} &= \hat{K}_{1} + i \hat{K}_{2} = \hat{a}_{1}^{\dagger} \hat{a}_{2}^{\dagger}, \\
        \hat{K}_{-} &= \hat{K}_{1} - i \hat{K}_{2} = \hat{a}_{1} \hat{a}_{2},
    \end{aligned}
\end{align}
It is important to stress that the algebra has a Casimir invariant \cite{Vourdas1990},
\begin{align}
    \begin{aligned}
        \hat{K}^{2} &=\hat{K}_{3}^{2}- \hat{K}_{1}^{2} - \hat{K}_{2}^{2}, \\
        &= \hat{K}_{3}^{2} - \frac{1}{2} \left( \hat{K}_{+} \hat{K}_{-} + \hat{K}_{-} \hat{K}_{+} \right), \\
        &=  \frac{1}{4}\left[ \left( \hat{a}_{1}^{\dagger}\hat{a}_{1} - \hat{a}_{2}^{\dagger}\hat{a}_{2} \right)^{2} - 1 \right],
    \end{aligned}
\end{align}
that commutes, 
\begin{align}
    \left[ \hat{K}^{2}, \hat{K}_{k} \right] = 0,
\end{align}
with all other elements of the algebra in any representation, $k = 1,2,3, \pm$.

It is possible to use a two-label basis that simultaneously diagonalize the Casimir $\hat{K}^{2}$ operator and one of the other operators \cite{Ban1993, Vourdas1990}, 
\begin{align}
    \begin{aligned}
        \hat{K}^{2} \vert k; m \rangle &= k (k-1) \, \vert k; m \rangle, \\
        \hat{K}_{3} \vert k; m \rangle &= (k+m) \, \vert k; m \rangle, \\
        \hat{K}_{+} \vert k; m \rangle &= \sqrt{(m + 1) (2k + m) }\, \vert k; m + 1\rangle, \\
        \hat{K}_{-} \vert k; m \rangle &= \sqrt{m (2k + m - 1) }\, \vert k; m - 1\rangle,
    \end{aligned}
\end{align}
with ladder operators $\hat{K}_{\pm}$.
The first label, the Bargmann parameter $k=(|n_{1}-n_{2}|+1)/2$, is related to the Casimir operator, that is related to the $\hat{J}_{3}$ operator of the $su(2)$ Lie algebra such that $\hat{K}^{2} = \hat{J}_{3}^{2} - 1/4$. 
There will be one subspace for the Bargmann parameter value $k=1/2$ and two for each other value $k = 1, 3/2, 5/2, \ldots$.
We will refer to the second label as the excitation number $ m = 0,1,2,\ldots$, such that each Bargmann subspace has infinite dimension \cite{Shaterzadeh2008}.
Thus, the connection between this complete orthonormal basis and the excitation number states, 
\begin{align}
    \begin{aligned}
    \vert k; m \rangle = \left\{ \begin{array}{ll} 
    \vert n_{1} = 2k + m - 1, n_{2} = m \rangle, \\
    \vert n_{1} = m, n_{2} = 2k + m - 1 \rangle, 
     \end{array} \right.
    \end{aligned}
\end{align}
is straightforward to make.
For lack of a better name, we will call these finite dimensional, discrete, complete orthonormal bases a Bargmann basis.

The $su(1,1)$ structure Lie algebra underlying the eigenvalue problem allows for the construction of a unitary squeezing, 
\begin{align}
    \begin{aligned}
        \hat{S}(\alpha) &= e^{\left( \alpha \hat{K}_{+} - \alpha^{\ast}  \hat{K}_{-} \right)}, \\
        &= e^{ i \theta \hat{K}_{z}} e^{ i 2 \vert \alpha \vert \hat{K}_{y}} e^{-i \theta \hat{K}_{z}},
    \end{aligned}
\end{align}
with complex coherent parameter $\alpha = \vert \alpha \vert e^{i \theta}$ that produces a squeezing and a rotation of the generalized dimensionless canonical pairs, 

\begin{align}
    \begin{aligned}
        \hat{S}^{\dagger}(\alpha) \, \hat{q}_{1} \, \hat{S}(\alpha) &= \hat{q}_{1} \cosh{\vert \alpha \vert} + \hat{q}_{2} \sinh{\vert \alpha \vert} \cos{\theta} + \hat{p}_{2} \sinh{\vert \alpha \vert} \sin{\theta}, \\
        \hat{S}^{\dagger}(\alpha) \, \hat{q}_{2} \, \hat{S}(\alpha) &= \hat{q}_{2} \cosh{\vert \alpha \vert} + \hat{q}_{1} \sinh{\vert \alpha \vert} \cos{\theta} + \hat{p}_{1} \sinh{\vert \alpha \vert} \sin{\theta}, \\
        \hat{S}^{\dagger}(\alpha) \, \hat{p}_{1} \, \hat{S}(\alpha) &= \hat{p}_{1} \cosh{\vert \alpha \vert} - \hat{p}_{2} \sinh{\vert \alpha \vert} \cos{\theta} + \hat{q}_{2} \sinh{\vert \alpha \vert} \sin{\theta}, \\
        \hat{S}^{\dagger}(\alpha) \, \hat{p}_{2} \, \hat{S}(\alpha) &= \hat{p}_{2} \cosh{\vert \alpha \vert} - \hat{p}_{1} \sinh{\vert \alpha \vert} \cos{\theta} + \hat{q}_{1} \sinh{\vert \alpha \vert} \sin{\theta}, 
    \end{aligned}
\end{align} 

proportional to the coherent parameter amplitude and phase in that order.
Sadly, it is too cumbersome to calculate the action of the unitary $su(1,1)$ rotation on a function of dimensionless configuration space as we did for the unitary displacement.
However, for a real coherent parameter, it is straightforward to see that the action in configuration space is a squeezing of the canonical variables.

The squeezing operator in terms of the creation and annihilation operators,  
\begin{align}
    \hat{S}(\alpha) &= e^{\left( \alpha \hat{a}_{1}^{\dagger} \hat{a}_{2}^{\dagger} - \alpha^{\ast}  \hat{a}_{1} \hat{a}_{2} \right)}, 
\end{align}
provides a squeezing with a complex phase, 

\begin{align}
    \begin{aligned}
        \hat{S}^{\dagger}(\alpha) \hat{a}_{1}  \hat{S}(\alpha) &= \hat{a}_{1} \cosh{\vert \alpha \vert} + \hat{a}_{2}^{\dagger} e^{i\theta} \sinh{\vert \alpha \vert}, \\
        \hat{S}^{\dagger}(\alpha) \hat{a}_{2}  \hat{S}(\alpha) &= \hat{a}_{2} \cosh{\vert \alpha \vert} + \hat{a}_{1}^{\dagger} e^{i\theta} \sinh{\vert \alpha \vert}, \\
        \hat{S}^{\dagger}(\alpha) \hat{a}_{1}^{\dagger}  \hat{S}(\alpha) &= \hat{a}_{1}^{\dagger} \cosh{\vert \alpha \vert} + \hat{a}_{2} e^{-i\theta} \sinh{\vert \alpha \vert},\\
        \hat{S}^{\dagger}(\alpha) \hat{a}_{2}^{\dagger}  \hat{S}(\alpha) &= \hat{a}_{2}^{\dagger} \cosh{\vert \alpha \vert} + \hat{a}_{1} e^{-i\theta} \sinh{\vert \alpha \vert},
    \end{aligned}
\end{align}

of the creation and annihilation operators where the amplitude of the complex parameter provides the rotation of the annihilation and creation operators. 

It is possible to squeeze any state in the Bargmann basis using  formulas for the exponential map of $su(1,1)$ operators in normal ordering \cite{Ban1993}. 
These states are known as generalized squeezed number states \cite{VillanuevaVergara2015}, 
\begin{align}
    \begin{aligned}
        \vert k; \alpha, m \rangle &= \hat{S}(\alpha) \vert k; m \rangle, \\
        &= \sqrt{\left( \begin{array}{c} 2k+m-1 \\ m \end{array} \right)} \, \mathrm{sech}^{2k} \vert \alpha \vert  \left( e^{ i \theta} \mathrm{tanh} \vert \alpha \vert \right)^{m} \times \\
        &\times \sum_{p=0}^{\infty} (-1)^{p} \sqrt{ \left( \begin{array}{c}2k+p-1 \\ p \end{array} \right)  } \,  \left( e^{ - i \theta } \mathrm{tanh} \vert \alpha \vert \right)^{p}   \,  \times \\
        & \times  \,_{2}\mathrm{F}_{1} \left(-p,-m; 2k; -\mathrm{csch}^{2} \vert \alpha \vert \right) \, \vert k; p \rangle, 
    \end{aligned}
\end{align}
such that in configuration space it takes a shape dependent on the representation used. 
However, it is possible to calculate the mean position and momentum for these generalized squeezed number states, 

\begin{align}
    \begin{aligned}
        \langle \hat{q}_{1} \rangle = \langle \hat{q}_{2} \rangle = \langle \hat{p}_{1} \rangle = \langle \hat{p}_{2} \rangle = 0,
    \end{aligned}
\end{align}
with variances, 
\begin{align}
    \begin{aligned}
        \sigma_{{q}_{1}}^{2}  &=\sigma_{{p}_{1}}^{2} = 2k \cosh^{2}{\vert \alpha \vert} + m \cosh{2\vert \alpha \vert} - \frac{1}{2},\\
        \sigma_{{q}_{2}}^{2} &= \sigma_{{p}_{2}}^{2} = 2k \sinh^{2}{\vert \alpha \vert} + m \cosh{2\vert \alpha \vert} + \frac{1}{2},
    \end{aligned}
\end{align}
that produce Heisenberg uncertainty relations, 

\begin{align}
    \begin{aligned}
        \sigma^{2}_{q_{1}} \sigma^{2}_{p_{1}} &= \left(2k \cosh^{2}{\vert \alpha \vert} + m \cosh{2\vert \alpha \vert} - \frac{1}{2} \right)^{2}, \\
        \sigma^{2}_{q_{2}} \sigma^{2}_{p_{2}} &= \left(2k \sinh^{2}{\vert \alpha \vert} + m \cosh{2\vert \alpha \vert} + \frac{1}{2} \right)^{2},
    \end{aligned}
\end{align}

that are not minimized for the squeezed vacuum state with $m=0$.
However, squeezed vacuum minimize combinations of the form $\sigma_{q_{1} \pm q_{2} }^{2} \sigma_{p_{1} \mp p_{2} }^{2}$ \cite{Bello2021}.

The generalized squeezed number states mean of the excitation number operators for the generalized coordinates, 
\begin{align}
    \begin{aligned}
        \langle \hat{n}_{1} \rangle &= (2k-1) \cosh^{2}{\vert \alpha \vert} + m \cosh{2 \vert \alpha \vert},\\
        \langle \hat{n}_{2} \rangle &= (2k-1) \sinh^{2}{\vert \alpha \vert} + m \cosh{2 \vert \alpha \vert},\\
    \end{aligned}
\end{align}
such that with variances, 
\begin{align}
    \begin{aligned}
        \sigma^{2}_{n_{1}} =\sigma^{2}_{n_{2}}  &= \frac{1}{4}(4km + m^{2} + 2k + m -1) \sinh^{2}{2\vert \alpha \vert},\\
    \end{aligned}
\end{align}

The mean expectation value for elements of the $su(1,1)$ algebra, 
\begin{align}
    \begin{aligned}
        \langle \hat{K}_{1} \rangle &= (k+m) \sinh{2\vert \alpha \vert} \cos{\theta}, \\
        \langle \hat{K}_{2} \rangle &= - (k+m) \sinh{2\vert \alpha \vert} \sin{\theta}, \\
        \langle \hat{K}_{3} \rangle &= (k+m) \cosh{2\vert \alpha \vert}, 
    \end{aligned}
\end{align}
that may be seen as describing the coordinates of a point on a hyperboloid of the form $x^{2} + y^{2} -z^{2} = -(k+m)^{2}$ with each subspace providing one sheet of the hyperboloid. 
The variances for the expected values of these operators, 

\begin{align}
    \begin{aligned}
        \sigma^{2}_{K_{1}} &= \frac{1}{2} \left(2km + m^{2} + k\right) \left(\cosh^{2}{2\vert \alpha \vert} \cos^{2}{\theta} + \sin^{2}{\theta}\right), \\
        \sigma^{2}_{K_{2}} &=  \frac{1}{2} \left(2km + m^{2} + k\right) \left(\cosh^{2}{2\vert \alpha \vert} \sin^{2}{\theta} + \cos^{2}{\theta}\right), \\
        \sigma^{2}_{K_{3}} &= \frac{1}{2} \left(2km + m^{2} + k\right) \sinh^{2}{2\vert \alpha \vert}, 
    \end{aligned}
\end{align}
lead to Heisenberg uncertainty relations that are too cumbersome to write here but that are minimized for squeezed vacuum states with $m=0$.

\subsection{Hermite-Gauss modes}

\begin{figure}
\includegraphics[width = 0.75 \linewidth]{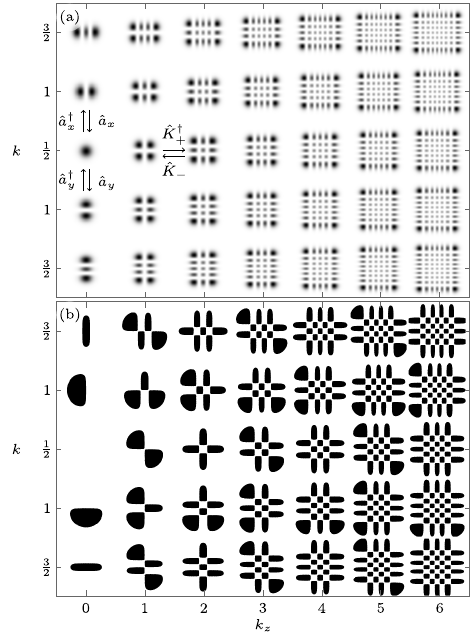}
\caption{(a) Irradiance and (b) phase distributions for Hermite--Gauss modes in two-mode $su(1,1)$ subspaces with constant $\langle \hat{K}^{2} \rangle = k (k-1)$. Each line is an infinite dimension subspace with Bargmann number $k = 1/2, 1, 3/2, \ldots$. Each Bargmann number has an horizontal and vertical configuration for its vacuum state except for $k=1/2$.} \label{fig:FigHGSU11}
\end{figure}

The Bagmann modes in Cartesian configuration space, 

\begin{align}
    \begin{aligned}
        \langle q_{x}, q_{y} \vert k; m \rangle &=~ \left\{ \begin{array}{l}
             \psi_{2k+m-1,m}(q_{x},q_{y}),\\
             \psi_{m,2k+m-1}(q_{x},q_{y}),
        \end{array} \right.   \\
        &=~ \dfrac{e^{-\frac{1}{2} (q_{x}^2+q_{y}^2)}}{\sqrt{ \pi  2^{2(k+m)-1}  (2k+m-1)! (m)! \, }}  \times \\
        & \times \left\{ \begin{array}{l}
              \mathrm{H}_{2k+m-1}(q_{x}) \mathrm{H}_{m}(q_{y} ), \\
              \mathrm{H}_{m}(q_{x}) \mathrm{H}_{2k+m-1}(q_{y} ),
        \end{array} \right.  
    \end{aligned}
\end{align}

answer to the Sturm-Liouville problem with eigenvalue $n = 2(k+m)-1$ that are double degenerate.
The underlying $su(1,1)$ Lie algebra structure suggest partitioning of the whole Hilbert space into subspaces of constant Casimir invariant, labelled by the Bargmann parameter $k$.
For each Bargmann parameter value with the exception of $k=1/2$, there are two configurations where the vacuum state shows either horizontal or vertical lobes.
Figures \ref{fig:FigHGSU11}(a) shows their irradiance distribution $\vert \psi_{2k+m-1,m}(q_{x},q_{y}) \vert^{2}$ and Fig. \ref{fig:FigHGSU11}(b) their phase distribution $\mathrm{arg} \left[ \psi_{2k+m-1,m}(q_{x},q_{y}) \right]$.
Each line in Fig. \ref{fig:FigHGSU11} is a subspace of constant Bargmann parameter $k$ with infinite elements. 
Under this underlying symmetry, the action of the $su(1,1)$ ladder operators moves us left or right in each line while the action of the creation and annihilation operators of the Heisenberg-Weyl Lie algebra moves us up and down in the state ladder, \ref{fig:FigHGSU11}(a).

\begin{figure}
\includegraphics[width = 0.75 \linewidth]{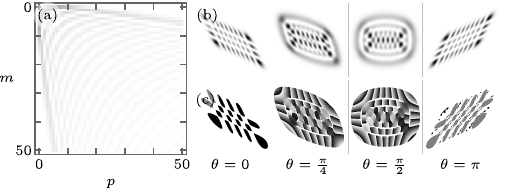}
\caption{(a) Probability $P_{k}(\alpha, m)$, (b) Irradiance $\vert \Psi_{\alpha, m} (q_{x}, q_{y}) \vert^{2}$, and (c) Phase $\mathrm{arg} \left[ \Psi_{\alpha, m} (q_{x}, q_{y}) \right]$ distributions for squeezed number states of a Hermite-Gauss mode with $k=3$, $m=3$, and complex coherent parameters $\alpha = \vert \alpha \vert e^{i \theta}$ with amplitude $\vert \alpha \vert = 1/2$} \label{fig:FigHGSqu}
\end{figure}

The squeezed number modes in Cartesian dimensionless configuration space take a non-intuitive form, 

\begin{align}
    \begin{aligned}
        \langle q_{x}, q_{y} \vert k; \alpha, m \rangle = &= \sqrt{\left( \begin{array}{c} 2k+m-1 \\ m \end{array} \right)} \, \mathrm{sech}^{2k} \vert \alpha \vert  \left( e^{ i \theta} \mathrm{tanh} \vert \alpha \vert \right)^{m} \times \\
        &\times \sum_{p=0}^{\infty} (-1)^{p} \sqrt{ \left( \begin{array}{c}2k+p-1 \\ p \end{array} \right)  } \,  \left( e^{ - i \theta } \mathrm{tanh} \vert \alpha \vert \right)^{p}   \,  \times \\
        & \times  \,_{2}\mathrm{F}_{1} \left(-p,-m; 2k; -\mathrm{csch}^{2} \vert \alpha \vert \right) \, \times \\
        & \times \, \dfrac{e^{-\frac{1}{2} (q_{x}^2+q_{y}^2)}}{\sqrt{ \pi  2^{2(k+p)-1}  (2k+p-1)! (p)! \, }}  \times \\
        & \times \left\{ \begin{array}{l}
              \mathrm{H}_{2k+p-1}(q_{x}) \mathrm{H}_{p}(q_{y} ), \\
              \mathrm{H}_{p}(q_{x}) \mathrm{H}_{2k+p-1}(q_{y} ),
        \end{array} \right.  
    \end{aligned}
\end{align}

with their irradiance centroid at the origin of configuration space. 
Figure \ref{fig:FigHGSqu}(a) shows the probability distribution for different initial Dicke states with excitation number parameter $m$.
Figure \ref{fig:FigHGSqu}(b) and Fig. \ref{fig:FigHGSqu}(c) show the irradiance distribution for a sample of these states.
A fixed real coherent parameter, $\theta = 0, \pi$, provides shearing og the original Dicke state depending on the , first and last column of Fig. \ref{fig:FigHGSqu}(b) and Fig. \ref{fig:FigHGSqu}(c).

\subsection{Laguerre-Gauss modes}
\begin{figure}
\includegraphics[width = 0.75 \linewidth]{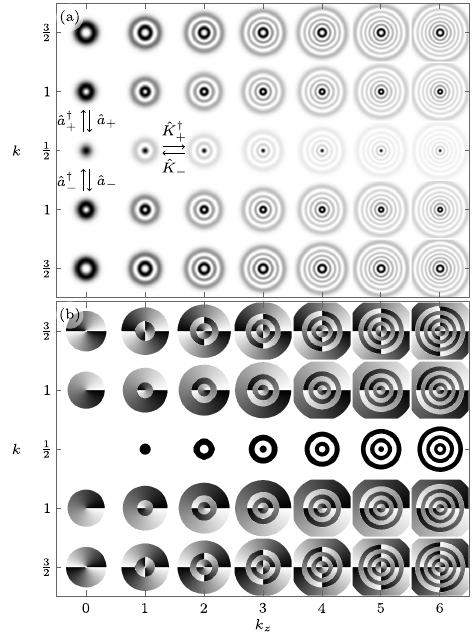}
\caption{(a) Irradiance and (b) phase distributions for Laguerre--Gauss modes in two-mode $su(1,1)$ subspaces with constant $\langle \hat{K}^{2} \rangle = k (k-1)$. 
Each line is an infinite dimension subspace with Bargmann number $k = 1/2, 1, 3/2, \ldots$. Each Bargmann number has a positive and negative topological charge configuration for its vacuum state except for $k=1/2$.} \label{fig:FigLGSU11}
\end{figure}

The Bargmann modes in Polar configuration space, 
\begin{align}
    \begin{aligned}
        \langle q_{\rho}, q_{\varphi} \vert k; m \rangle &=~
             \psi_{m,\pm (2k+m-1)}(q_{\rho},q_{\varphi}), \\
        &=\frac{1}{\sqrt{\pi }} (-1)^{m} \sqrt{\frac{ m!}{ \left( m+ 2k -1 \right)!}} \times\\
        &~ \times \, q_{\rho}^{(2k - 1)} e^{ - \frac{1}{2} q_{\rho}^{2}} \mathrm{L}_{m}^{( 2k -1) } (q_{\rho}^{2}) e^{\pm i (2k - 1) q_{\varphi}} ,  
    \end{aligned}
\end{align}
answer to the Sturm-Liouville problem with eigenvalue $n = 2(k+m)-1$ that are double degenerate.
The underlying $su(1,1)$ Lie algebra structure suggest partitioning of the whole Hilbert space into subspaces of constant Casimir invariant, labelled by the Bargmann parameter $k$ that share the same topological charge $\ell = \pm (2k - 1)$.
For each Bargmann parameter value with the exception of $k=1/2$, there are two configurations where the vacuum state shows positive or negative topological charge.
Figures \ref{fig:FigLGSU11}(a) shows their irradiance distribution $\vert \psi_{m,\pm(2k+m-1)}(q_{\rho},q_{\varphi}) \vert^{2}$ and Fig. \ref{fig:FigLGSU11}(b) their phase distribution $\mathrm{arg} \left[ \psi_{m,\pm(2k+m-1)}(q_{\rho},q_{\varphi}) \right]$.
Each line in Fig. \ref{fig:FigLGSU11} is a subspace of constant Bargmann parameter $k$ with infinite number of elements. 
Under this underlying symmetry, the action of the $su(1,1)$ ladder operators moves us left or right in each line while the action of the creation and annihilation operators of the Heisenberg-Weyl Lie algebra moves us up and down in the state ladder,  \ref{fig:FigLGSU11}(a).

\begin{figure}
\includegraphics[width = 0.75 \linewidth]{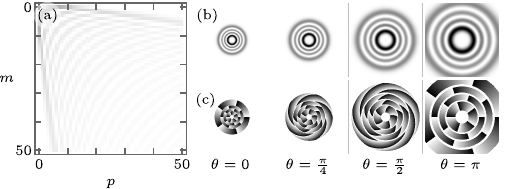}
\caption{(a) Probability $P_{k}(\alpha, m)$, (b) Irradiance $\vert \Psi_{\alpha, m} (q_{x}, q_{y}) \vert^{2}$, and (c) Phase $\mathrm{arg} \left[ \Psi_{\alpha, m} (q_{x}, q_{y}) \right]$ distributions for squeezed number states of a Laguerre-Gauss mode with $k=3$, $m=3$,  and complex coherent parameters $\alpha = \vert \alpha \vert e^{i \theta}$ with amplitude $\vert \alpha \vert = 1/2$ and phase values $\theta = 0, \pi/4, \pi/2, \pi$.} \label{fig:FigLGSqu}
\end{figure}

The squeezed number modes in Polar dimensionless configuration space take a non-intuitive form, 

\begin{align}
    \begin{aligned}
        \langle q_{\rho}, q_{\varphi} \vert k; \alpha, m \rangle &= \sqrt{\left( \begin{array}{c} 2k+m-1 \\ m \end{array} \right)} \, \mathrm{sech}^{2k} \vert \alpha \vert  \left( e^{ i \theta} \mathrm{tanh} \vert \alpha \vert \right)^{m} \times \\
        &\times \sum_{p=0}^{\infty} \sqrt{ \left( \begin{array}{c}2k+p-1 \\ p \end{array} \right)  } \,  \left( e^{ - i \theta } \mathrm{tanh} \vert \alpha \vert \right)^{p}   \,  \times \\
        & \times  \,_{2}\mathrm{F}_{1} \left(-p,-m; 2k; -\mathrm{csch}^{2} \vert \alpha \vert \right)  \sqrt{\frac{ p!}{\left( p + 2k -1 \right)!}} \times\\
        &~ \times \, \frac{1}{\sqrt{\pi}}q_{\rho}^{(2k - 1)} e^{ - \frac{1}{2} q_{\rho}^{2}} \mathrm{L}_{p}^{( 2k -1) } (q_{\rho}^{2}) e^{\pm i (2k - 1) q_{\varphi}}, 
    \end{aligned}
\end{align}
with their irradiance centroid at the origin of configuration space. 
Figure \ref{fig:FigLGSqu}(a) shows the probability distribution for different initial Dicke states with excitation number parameter $m$.
Figure \ref{fig:FigLGSqu}(b) and Fig. \ref{fig:FigLGSqu}(c) show the irradiance distribution for a sample of these states.
The coherent parameter phase controls the scaling and shearing of the irradiance and phase distributions, in that order.
In the phase range $\theta \in \left[0, \pi\right]$ the irradiance pattern scales up and down for the range $\theta \in (\pi, 2 \pi)$ while the phase will shear counter-clockwise and clockwise in that order.

\section{Experimental details} \label{sec:S7}

Consider the mode $\psi(\vect{r}_\perp) = \left|\psi(\vect{r}_\perp)\right| e^{i \mathrm{Arg}[\psi(\vect{r}_\perp)]}$, that we aim to experimentally reproduce in the lab.  Various techniques exist for generating optical modes that require complex modulation, such as those developed in Refs.\ \cite{Kirk1971, Davis1999, Arrizon2007, Bolduc2013}.  To achieve this, we can use a phase-only Spatial Light Modulator (SLM) to display a digital hologram and create the desired optical mode.  

\begin{figure}[htp!]
	\centering
	\includegraphics[width= 0.5 \linewidth]{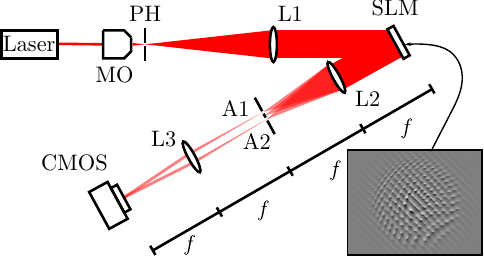}
	\caption{Experimental setup.  Laser: He--Ne source; MO: Microscope Objective; PH: Pinhole; L1--L3: lenses; A1--A2: Apertures; SLM: Spatial Light Modulator; CMOS: sensor.  The inset shows an example of the digital hologram displayed on the screen of the Spatial Light Modulator.} 
	\label{fig:FigSetup}
\end{figure}

Figure \ref{fig:FigSetup} depicts our experimental setup to observe the classical analogues of the generalized coherent states.  We used a low--power HeNe laser source ($\lambda = 633$ nm), which was expanded and collimated to illuminate the screen of a Spatial Light Modulator (SLM Holoeye Pluto VIS).  The SLM displayed a digital hologram \cite{Forbes2016} for the classical versions of our coherent states.  A system of two lenses (L2 and L3) and precise filtering via A1 and A2, enabled us to experimentally study the aforementioned modes.  The intensity of the optical field was recorded in the back focal plane of L3 using a CMOS sensor (Nikon D3100).

To measure the transverse phase of the generated beam, we employed the four-step technique \cite{Gaasvik2003}.  Consider the interferogram between the mode $\psi_{n_1, n_2}(\vect{r}_\perp)$ and a reference wave $U_r(\vect{r}_\perp)$, which can be expressed as
\begin{align*}
	I_j(\vect{r}_\perp) &= \left|\psi_{n_1, n_2}(\vect{r}_\perp) + e^{i\beta_j} U_r(\vect{r}_\perp) \right|^2,\\
	&= |\psi_{n_1, n_2}(\vect{r}_\perp)|^2 + |U_r(\vect{r}_\perp)|^2 + \\
	&\hspace{13pt}2 |\psi_{n_1, n_2}(\vect{r}_\perp)| |U_r(\vect{r}_\perp)| \:\cos\left[\mathrm{arg}[\psi_{n_1, n_2}(\vect{r}_\perp)] + \beta_j\right],
\end{align*}
where $\beta_j = (j-1)\pi/2$ with $j = 1,2,3,4$.  It is straightforward to see that the transverse phase distribution can be computed by
\begin{align*}
	\mathrm{arg}[\psi_{n_1, n_2}(\vect{r}_\perp)] = \mathrm{atan}\left[\frac{I_4(\vect{r}_\perp) - I_2(\vect{r}_\perp)}{I_1(\vect{r}_\perp) - I_3(\vect{r}_\perp)}\right].
\end{align*}

\begin{figure}[htp!]
	\centering
	\includegraphics[width= 0.75 \linewidth]{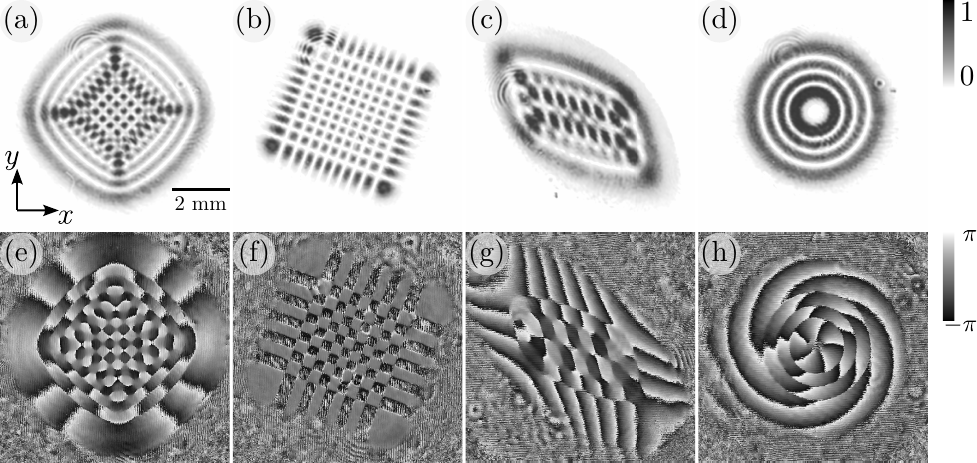}
	\caption{Experimental results of the generalized quantum coherent states.  The top row shows the intensity, while the bottom row their corresponding phases.  (a), (e) Generalized spin coherent states of a Hermite--Gauss mode with $j=10$, $m=0$, $|\alpha|=\pi/4$ and $\theta=\pi/4$.  (b), (f) Generalized spin coherent states of a Laguerre--Gauss mode with $j=10$, $m=0$, $|\alpha|=\pi/4$ and $\theta=\pi/4$. (c), (g) Squeezed number states of a Hermite--Gauss mode with $k=3$, $m=3$, $|\alpha|=1/2$ and $\theta=\pi/4$. (d), (h) Squeezed number states of a Laguerre--Gauss mode with $k=3$, $m=3$, $|\alpha|=1/2$ and $\theta=\pi/4$.  In all cases $w_0=0.75$ mm.}
	\label{fig:FigExp}
\end{figure}

A multiplexed hologram \cite{Forbes2016, Rosales-Guzman2017} was encoded on the SLM screen, where two beams were simultaneously generated.  Specifically, the desired optical mode was directed along the wave vector $\vect{k}_1$, whereas the reference beam (with a phase shift) was propagated in the $\vect{k}_2$ direction:
\begin{align}
    G(\vect{r}_\perp, z=0) = e^{i\vect{k}_1\cdot\vect{r}} \psi_{n_1, n_2}(\vect{r}_\perp) + e^{i\vect{k}_2\cdot\vect{r}} e^{i\beta_j} U_r(\vect{r}_\perp).
\end{align}
Proper filtering (A1 and A2) allowed both beams to propagate and interfere in the sensor plane.  It is important to remark, that the required phase shifts were embedded within the digital hologram, thus eliminating the need for movable optical parts.  Figure \ref{fig:FigExp} demonstrates the experimental validation of our generalized quantum coherent states. The intensity distribution is shown in the top row, while the transverse phase is depicted in the bottom row. The first and second columns show the generalized spin coherent states of Hermite- and Laguerre-Gauss modes, respectively, with parameters $j=10$, $m=0$, $|\alpha|=\pi/4$ and $\theta=\pi/4$.  The last two columns present the squeezed number states of Hermite- and Laguerre-Gauss modes, respectively, with parameters $k=3$, $m=3$, $|\alpha|=1/2$ and $\theta=\pi/4$.  In the four cases the beam waist was set to 0.75 mm.

\section{Conclusion} \label{sec:S8}

We used the spatial degree of freedom of light to construct classical optical analogues of generalized coherent states, using Hermite- and Laguerre-Gauss modes as building blocks. 
Our optical analogues preserve the statistical properties of their quantum counterparts, encoded in their amplitude and phase distributions. 

We explored the optical analogues of displaced (nonlinear coherent), rotated (generalized spin coherent), and squeezed number states.
The practical implication of our classical analogues are significant, particularly in the fields of metrology and sensing, where coherent and squeezed states provide an advantage. 
Additionally, including the polarization degree of freedom opens the possibility of implementing optical analogues to more complex measurement and sensing protocols that leverage correlations between polarization and spatial degrees of freedom.

Looking forward, we anticipate the exploration of the quantum state zoo that remains unexplored within the classical realm, such as intelligent states or displaced-squeezed states. 
Our approach demonstrates that spatial light modes enable the simulation of nonclassical states relevant for quantum technologies. Given the inherent robustness of classical light beams against noise and decoherence, the spatial modes discussed here hold great potential for developing robust, quantum-inspired sensing of photosensitive materials.
We hope our findings inspire continued exploration and development in this interesting area of study.

\section*{Funding}
O.S.M.L.: U.S. Department of Energy (DOE), Office of
Science (SC), Program of Nuclear Physics, NP-QIS grant: DE-SC0023694.

\section*{Acknowledgment}
M.~P.~M.~R. and B.~M.~R.~L. acknowledge fruitful discussions with Luis Miguel Nieto Calzada.

\section*{Disclosures} 
The authors declare no conflicts of interest.

\section*{Data availability} 
Data underlying the results presented in this paper may be obtained from the authors upon reasonable request.

\section*{References} 

%

\end{document}